\documentclass[sn-mathphys-num]{sn-jnl}% Math and Physical Sciences 

\usepackage{graphicx}%
\usepackage{multirow}%
\usepackage{amsmath,amssymb,amsfonts}%
\usepackage{amsthm}%
\usepackage{mathrsfs}%
\usepackage[title]{appendix}%
\usepackage{xcolor}%
\usepackage{textcomp}%
\usepackage{manyfoot}%
\usepackage{booktabs}%
\usepackage{algorithm}%
\usepackage{algorithmicx}%
\usepackage{algpseudocode}%
\usepackage{listings}%
\usepackage{dcolumn}
\usepackage{bm}
\usepackage{hyperref}
\usepackage{eucal}
\usepackage{subcaption}

\newcommand\numberthis{\addtocounter{equation}{1}\tag{\theequation}}
\usepackage[capitalise]{cleveref}

\theoremstyle{thmstyleone}%
\theoremstyle{thmstyletwo}%

\theoremstyle{thmstylethree}%

\newcommand{\firstphi}{\phi^{(1)}}
\newcommand{\firstpsi}{\psi^{(1)}}
\newcommand{\firstw}{w^{(1)}}
\newcommand{\firsth}{h^{(1)}}
\newcommand{\firstsigma}{\sigma^{(1)}}
\newcommand{\tensorh}{h^{i(2)}{_{j}}}

\newcommand{\dfirstphi}{\phi^{(1)'}}
\newcommand{\dfirstpsi }{\psi^{(1)'}}
\newcommand{\dfirstw}{w^{(1)'}}
\newcommand{\dfirsth}{h^{(1)'}}
\newcommand{\dfirstsigma}{\sigma^{(1)'}}

\newcommand{\ddfirstphi}{\phi^{(1)''}}
\newcommand{\ddfirstpsi}{\psi^{(1)''}}

\newcommand{\ddfirsth}{h^{(1)''}}

\newcommand{\ddtensorh}{h^{i(2)''}{_{j}}}
\newcommand{\dtensorh}{h^{i(2)'}{_{j}}}

\begin{document}

\title[SIGW generic gauge]{Gauge-dependence of Scalar Induced Gravitational Waves}

\author[1,2]{\fnm{Anjali Abirami} \sur{Kugarajh}}\email{anjali.kugarajh@gssi.it}

\affil[1]{\orgname{Gran Sasso Science Institute}, \orgaddress{\street{Viale F. Crispi 7}, \city{L'Aquila}, \postcode{I-67100}, \country{Italy}}}

\affil[2]{\orgdiv{Laboratori Nazionali del Gran Sasso}, \orgname{INFN}, \orgaddress{ \city{Assergi}, \postcode{I-67100}, \country{L'Aquila}}}

\abstract{In this review we look into the gauge-dependence of scalar-induced gravitational waves (SIGWs) that are second-order tensors produced by first-order scalar modes. The method includes deriving the background, first- and second-order Einstein field equations without imposing a gauge. We address the gauge-invariant approach and study the source-term of SIGWs in three different gauges: synchronous, Poisson, and uniform curvature gauge. We find that numerically computed kernels evaluated in a radiation epoch in all three gauges behave closely with minimal discrepancy. As expected, when going to sub-horizon modes, \( k\tau \gg 1 \), the discrepancy decreases and the behavior converges, suggesting that SIGWs can be treated as gauge-invariant observables in this regime.}


\maketitle

\section{\label{sec:introduction}Introduction}

Inflationary cosmology leads to the generation of quantum vacuum fluctuations in the metric, giving rise to scalar perturbations—observable today as density fluctuations—and tensor perturbations, which manifest as gravitational waves \cite{Guth:1982ec,Wands:2008tv}. These primordial gravitational waves (GWs) are expected to form a stochastic gravitational wave background (SGWB), reflecting their quantum origin. Several early Universe processes act as sources of GWs \cite{Maggiore:2007ulw,Maggiore:2018sht,Caprini:2018mtu,Guzzetti:2016mkm,Bartolo:2019oiq, Bartolo:2019yeu}, one of those from scalar-induced gravitational waves (SIGWs) \cite{10.1143/PTP.37.831, PhysRevD.47.1311, Matarrese:1993zf, Pantano:1995wvf, Matarrese:1997ay, Ananda:2006af}. These are observed as a second-order source of GWs which arise from the nonlinear coupling of large scalar fluctuations at horizon re-entry \cite{Acquaviva:2002ud,Mollerach:2003nq} \footnote{Here and throughout, we refer to ‘second-order’ as perturbations of order \( \mathcal{O}(\epsilon^2) \) arising from products of first-order scalar quantities \( \mathcal{O}(\epsilon) \).}. The frequency range of SIGWs produced in the radiation dominated era fall within current and future generation ground- and space-based detectors, for example Laser Interferometer Space Antenna (LISA) \cite{LISACosmologyWorkingGroup:2022jok, Caprini:2019pxz} and Pulsar Timing Array (PTA) \cite{NANOGrav:2023gor, Figueroa:2023zhu}. 

A perturbative approach is used to study SIGWs, where small inhomogeneous perturbations on a homogeneous and isotropic Friedmann-Roberston-Walker (FRW) background are considered \cite{Lifshitz:1945du}. This splitting of an unperturbed background and perturbed system introduces the complication in gauge choice. There is no preferred choice of coordinate in relativistic perturbation theory, allowing one to work in the appropriate gauge best suited for the problem. It is well known that at linear-order, scalar modes are gauge-dependent while tensor modes are gauge-independent, i.e. do not change under coordinate transformation \cite{Malik:2008im}. However due to the non-linear feature of the source, the second-order tensor modes that are sourced by these first-order scalar modes must also depend on the gauge conditions \cite{Arroja:2009sh}. In previous studies of ``scalar-induced” tensor modes, a gauge condition is usually imposed beforehand, usually the zero-shear gauge, also knows as longitudinal or Poisson gauge.

Physically, this raises the question: \textit{do SIGWs have any residual gauge dependence at second order?} Detailed calculations has been done in previous studies presenting the gauge-dependence of induced GWs by studying different gauges \cite{Hwang:2017oxa, Gong:2019mui, Tomikawa:2019tvi}. Some works aim to bypass this issue by constructing a gauge-invariant formalism for the second-order tensor modes by the addition of terms with the intention to reduce the gauge dependence \cite{Wang:2019zhj, Chang:2020iji, Chang:2020mky, Chang:2020tji}. However, these constructions can blur the connection between theoretical quantities and actual observables \cite{DeLuca:2019ufz, Domenech:2020xin}. They mention that there is not much difference between choosing particular gauge from the beginning and working with specific gauge-invariant combination. 

Instead, the second approach in tackling the gauge issue is by defining a gauge suitable to best describe the observable \cite{DeLuca:2019ufz, Inomata:2019yww, Domenech:2020xin}. In \cite{DeLuca:2019ufz}, they argue that the suitable gauge is one in which the coordinates are fixed according to the positions of the interferometer mirrors. This is known as the synchronous gauge, or \textit{TT gauge}. \cite{DeLuca:2019ufz} show that the observable GW spectrum of GWs in the TT gauge and Poisson gauge coincide. It is important to note that scalar-induced tensor perturbations can be interpreted as freely propagating gravitational waves—and thus effectively treated as linear, first-order metric perturbations—only once the source term becomes inactive and the modes are well inside the horizon, in the deep sub-horizon regime. In this context, the gauge dependence associated with the perturbations no longer persists, as the gravitational waves evolve independently. A similar argument was made by \cite{Domenech:2020xin}, in which they show that the scalar-induced gravitational waves behave similar when calculated in gauges which are ``well-behaved" on small scales. The authors point out that the Newtonian gauge is suitable both for the calculation and the physical interpretation. Further mentioning other gauges, among them the synchronous gauge, to also be ``well-behaved" in small scales in a universe dominated by perfect fluid with constant equations of state.

Despite significant progress in previous studies, certain limitations persist. In particular, many approaches impose gauge conditions at an early stage, which can hide leftover gauge effects. While gauge-invariant formulations offer a rigorous framework, they can still present challenges in relating theoretical constructs to measurable quantities. In this review, we present a general framework for deriving the equation of motion and source term for SIGWs without imposing a specific gauge choice. This approach provides a unified and practical framework for evaluating the source term. We examine its behavior across three commonly used gauges—synchronous, Poisson, and uniform curvature—to illustrate how gauge choices influence the resulting expressions. By conducting both analytical and numerical comparisons of the corresponding kernels, we systematically quantify gauge-related discrepancies and clarify the degree to which SIGWs can be considered gauge-invariant at the level of observables.

In \cref{sec:perturbation_theory,sec:einstein_field_eq}, we introduce the general framework for cosmological perturbations of the metric and then derive the perturbed Einstein field equations. In \cref{subsec:second_order_eqs}, the source term is presented without imposing a gauge. Further on the gauge transformation and gauge-invariant variables are presented in \cref{sec:gauge_first_order}. We work with the approaches mentioned to show the behaviour of the scalar-induced gravitational waves in different gauges. In \cref{sec:source_term_gauges,sec:gauges_kernel}, we study the source-term of the SIGWs and the subsequent kernel in three different gauges.

In the following calculations, we will be using natural units, $c = \hbar = 1$ and Planck mass $M_{p}^{-2} = 8\pi G = \kappa^{2}$. In addition, we are considering conformal time, $\tau$, where $a(\tau) d\tau = dt$. To denote four-dimensional spacetime indices, we have used Greek indices, $\mu, \nu$. While Roman indices are used to denote spatial indices $i, j$. The total space-time metric has signature $(-,+,+,+)$. 

\section{\label{sec:perturbation_theory}Cosmological perturbation of the metric}

This section introduces the perturbative framework used to study small fluctuations around the homogeneous and isotropic FRW universe. The decomposition into scalar, vector, and tensor modes allows us to trace how different physical effects—such as density fluctuations and gravitational waves—emerge from different types of perturbations. Understanding this structure is fundamental before identifying which modes source second-order gravitational waves.

Any tensorial quantity \(A\) can be decomposed into background and inhomogeneous parts. This decomposition is performed in a Friedmann–Lemaître–Robertson–Walker (FLRW) cosmological spacetime, employing a foliation defined by the conformal time \(\tau\), such that the hypersurfaces of constant \(\tau\) are spatially homogeneous and isotropic. This standard background provides the basis for the perturbative expansion used throughout the paper. The decomposition follows
\begin{equation}
    \mathbf{A} = \mathbf{A}^{(0)}(\tau) + \delta^{(r)}\mathbf{A}(\tau, x^{i}).  
\end{equation}
The background quantity is $\mathbf{A}^{(0)}$ and only has time-dependence. The perturbed quantity up to $r$-th order, $\delta^{(r)}\mathbf{A}$, is defined as 
\begin{equation}
	\delta^{(r)}\mathbf{A} = \sum_{r=1}^{\infty}\frac{1}{r!}\delta^{(r)}\mathbf{A},
\end{equation}
where it has both time- and spatial-dependence. The metric tensor, $g_{\mu\nu}$, up to the second order is
\begin{equation}
\label{metric_tensor_decomposed}
    g_{\mu\nu} = g_{\mu\nu}^{(0)} + \delta^{(1)} g_{\mu\nu} + \frac{1}{2}\delta^{(2)}g_{\mu\nu}, 
\end{equation}
where $\delta g_{\mu\nu}$ and $\delta^{(2)} g_{\mu\nu}$ is the perturbed metric in first and second order. Using a method known as (3+1) decomposition, also known as scalar--vector--tensor (SVT) decomposition, 3-dimensional quantities are split into scalar, vector and tensor components corresponding to the transformational behavior on spatial hypersurfaces. The scalar, vector and tensor modes are defined as follows \cite{Matarrese:1997ay, Malik:2008im}:
\begin{itemize}
    \item \textbf{Scalar perturbations:} \( \phi^{(r)} \), \( \psi^{(r)} \), \( w^{(r)} \), and \( h^{(r)} \),
    \item \textbf{Vector perturbations:} transverse \( w^{(r)}_i \), \( h^{(r)}_i \),
    \item \textbf{Tensor perturbations:} transverse-traceless \( h^{(r)}_{ij} \).
\end{itemize}
Throughout this work, we denote the perturbation order with a superscript in parentheses, e.g., \( \phi^{(1)} \), \( h^{(2)}_{ij} \). The components of the perturbed spatially flat Robertson - Walker metric can be written as
\begin{align*}
\label{metric_tensor}
    g_{00} & = -a^{2}(\tau) \left(1 + 2\sum_{r=1}^{\infty}\frac{1}{r!}\phi^{(r)}(\vec{x},\tau)\right), \\
    g_{0i} & = a^{2}(\tau)\sum_{r=1}^{\infty}\frac{1}{r!}\hat{w}{_{i}}^{(r)}(\vec{x},\tau), \\
    g_{ij} & = a^{2}(\tau)\left\{\left[1-2\left(\sum_{r=1}^{\infty}\frac{1}{r!}\psi^{(r)}(\vec{x},\tau)\right)\right]\delta_{ij}+\sum_{r=1}^{\infty}\frac{1}{r!}\hat{h}_{ij}^{(r)}(\vec{x},\tau)\right\}, \numberthis
\end{align*}
where the $0-i$ and $i-j$ components can be further decomposed, 
\begin{equation}
\label{vector_decompose}
    \hat{w}{_{i}}^{(r)} = \partial_{i}w^{(r)} + w_{i}^{(r)},
\end{equation}
\begin{equation}
\label{tensor_decompose}
    \hat{h}_{ij}^{(r)} = D_{ij}h^{(r)} + \partial_{i}h_{j}^{(r)} + \partial_{j}h_{i}^{(r)} + h_{ij}^{(r)}.
\end{equation}
The vector perturbations are transverse, i.e. $\partial^{i}w_{i}^{(r)} = \partial^{i}h_{i}^{(r)} = 0$, and $h_{ij}^{(r)}$ is a symmetric transverse and trace-free tensor, i.e. $\partial^{i}h_{ij}^{(r)} = h_{i}^{i(r)} = 0$. Finally, $D_{ij} = \partial_{i}\partial_{j} - \frac{1}{3}\nabla^{2}\delta_{ij}$ is the trace-free operator.

In order to study second-order tensor modes, the metric \cref{metric_tensor} can be simplified further with the following assumptions; (1) first-order vector perturbations have decreasing amplitude, thus being able to neglect $w^{(1)}{_{i}}$ and $h^{(1)}{_{i}}$, (2) first-order tensors give negligible contribution to second-order perturbations, hence neglecting $h_{ij}^{(1)}$. On the other hand, we will keep the second-order vector and tensor perturbations since they can be generated by first-order scalar perturbations, i.e. the tensor modes being ``scalar-induced'' gravitational waves. With these assumptions the metric, \cref{metric_tensor}, can be re-written as 
\begin{center}
\begin{align*}
\label{metric_tensor_second_order}
    g_{00} & = -a^{2}(\tau) \left(1 + 2\firstphi + \phi^{(2)}\right), \\ 
    g_{0i} & = a^{2}(\tau)\left(\partial_{i}\firstw + \frac{1}{2}\partial_{i}w^{(2)} + \frac{1}{2}w_{i}^{(2)}\right), \\
    g_{ij} & = a^{2}(\tau) \left[\left(1 - 2\firstpsi - \psi^{(2)}\right)\delta_{ij} + D_{ij}\left(\firsth + \frac{1}{2}h^{(2)}\right) \right. \\
    & \left. + \frac{1}{2}\left(\partial_{i}h_{j}^{(2)} + \partial_{j}h_{i}^{(2)} + h_{ij}^{(2)}\right)\right]. \numberthis
\end{align*} 
\end{center}
The contra-variant metric tensor can be obtained using the following relation, $g_{\mu\nu}g^{\nu\lambda} = \delta^{\lambda}_{\mu}$, where
\begin{center}
\begin{align*}
\label{contravariant_metric_tensor_second_order}
    g^{00} & = -a^{-2}(\tau) \left(1 - 2\firstphi - \phi^{(2)} + 4 \left(\firstphi\right)^{2} - \partial^{i}\firstw\partial_{i}\firstw\right), \\ 
    g^{0i} & = a^{-2}(\tau)\left[\partial^{i}\firstw + \frac{1}{2}\left(\partial^{i}w^{(2)} + w^{i(2)}\right) + 2\left(\firstpsi - \firstphi\right)\partial^{i}\firstw \right.\\
    & \left. - \partial^{k}\firstw D^{i}{_{k}}\firsth\right], \\
    g^{ij} & = a^{-2}(\tau)\left[\left(1 + 2\firstpsi + \psi^{(2)} + 4 \left(\firstpsi\right)^{2}\right)\delta^{ij} + D^{ij}\left(\firsth + \frac{1}{2}h^{(2)}\right) \right. \\
    & \left. - \frac{1}{2}\left(\partial^{i}h^{j(2)} + \partial^{j}h^{i(2)} + h^{ij(2)}\right) - \partial^{i}\firstw\partial^{j}\firstw \right. \\
    & \left. - 4\firstpsi D^{ij}\firsth + D^{ik}\firsth D^{j}{_{k}}\firsth\right]. \numberthis
\end{align*}  
\end{center}

\section{\label{sec:einstein_field_eq}Einstein field equations}

To determine the dynamics of the perturbations introduced earlier, we derive the Einstein field equations order by order. The key focus is on how scalar perturbations evolve and interact at second order to source tensor modes—i.e., gravitational waves. This derivation lays the groundwork for computing the source term of SIGWs. The Einstein field equation is defined as
\begin{equation}
	 G^{\mu}{_{\nu}}  = R^{\mu}{_{\nu}} - \frac{1}{2} g^{\mu}{_{\nu}}R=\kappa^{2}T^{\mu}{_{\nu}}.
\end{equation}
In the following sub-sections we study, background, first- and second-order Einsteins equations in a generic gauge. The components in order to find the field equations is presented in Appendix \ref{ricci_tensor}, \ref{ricci_scalar}, \ref{einstein_tensor}, \ref{energy_momentum_tensor}. 

\subsection{\label{subsec:bckg_eqs}Background Einsteins equations}

Using the background zeroth-order Einstein equation, the following terms are defined
\begin{align*}
\label{background_H}
    \mathcal{H}^{2} & = \frac{\kappa^{2} a^{2}}{3}\bar{\rho}, \\
    \mathcal{H}' & = -\frac{\kappa^{2} a^{2}}{6}\left(\bar{\rho} + 3\bar{P}\right), \numberthis
\end{align*}
where $\bar{P}$ and $\bar{\rho}$ are the background quantities for pressure and energy-density and 
$\mathcal{H} = \frac{a'}{a}$ is the conformal Hubble parameter\footnote{The prime notation indicates derivatives with respect to conformal time $\tau$.}.  

\subsection{\label{subsec:first_order_eqs}First order Einsteins equations}

The first-order Einstein field equations are used to find expressions for first-order perturbed matter quantities. Using the time-time component, the expression for first-order energy density is found
\begin{align*}
\label{00_component_evolution}
    G^{(1)0}{_{0}} & = \kappa^{2} T^{(1)0}{_{0}}, \\
    a^{-2}\left[6\mathcal{H}^{2}\firstphi + 6\mathcal{H}\dfirstpsi + 2\mathcal{H}\nabla^{2}\firstw - 2\nabla^{2}\firstpsi - \frac{1}{2}\partial_{k}\partial^{i} D^{k}{_{i}}\firsth\right] & = -\kappa^{2}\rho^{(1)}, \numberthis 
\end{align*}
The time-space component provides the first-order 4-velocity expression
\begin{align*}
\label{0i_component_evolution}
    G^{(1)0}{_i} & = \kappa^{2} T^{(1)0}{_{i}}, \\
    a^{-2}\left(-2\mathcal{H}\partial_{i}\firstphi - 2\partial_{i}\dfirstpsi - \frac{1}{2}\partial_{k} D^{k}{_{i}}\dfirsth\right) & = \kappa^{2}\left(\bar{\rho} + \bar{P}\right)\left(\partial_{i}\firstw + v^{(1)}{_{i}}\right), \numberthis
\end{align*}
We can find the trace and trace-free part of spatial component, $G^{(1)i}{_j} = \kappa^{2}T^{(1)i}{_j}$ ,
\begin{align*}
\label{ij_component_trace}
     - a^{-2}&\left(2\mathcal{H}\dfirstphi  + \left[4\mathcal{H}'+ 2\mathcal{H}^{2}\right]\firstphi + 4\mathcal{H}\dfirstpsi + 2\ddfirstpsi\right. \\
    & \left.  + \frac{2}{3}\nabla^{2}\left(\firstphi - \firstpsi + 2\mathcal{H}\firstw + \dfirstw\right)\right)  = \kappa^{2}P^{(1)}, \numberthis
\end{align*}
\begin{equation}
\label{ij_component_traceless}
    -a^{-2}\left(\partial^{i}\partial_{j} - \frac{1}{3}\nabla^{2}\delta^{i}{_{j}}\right)\left(\firstphi - \firstpsi + 2\mathcal{H}\firstw + \dfirstw\right) = \kappa^{2}a^{-2}\pi^{(1)i}{_{j}},
\end{equation}
Here \( \rho^{(1)} \), \( v^{(1)}_i \), and \( \pi^{(1)}_{ij} \) are the first-order scalar, vector, and tensor components of the energy-momentum tensor, respectively.

\subsection{\label{subsec:second_order_eqs}Second-order Einstein field equation and source term}

For understanding the source-term of scalar-induced gravitational waves, we will be studying the spatial component of second-order Einstein tensor, specifically the transverse, trace-free part. We do this by first applying the projection tensor $P^{li}_{jm}$ to the field equation \cite{Carbone:2004iv}, i.e. $P^{li}_{jm}G^{(2)m}{_{l}} = \kappa^{2}P^{li}_{jm}T^{(2)m}{_{l}}$. Accordingly, the second-order, scalar, vector and tensor modes are removed. The projected spatial Einstein tensor, \cref{non_diagonal_second_order_G}, and energy-momentum tensor, \cref{energy-momentum_second}, is presented in Appendix \ref{projected}. 
Using the relation for the equation of state, $\omega=\bar{P}/\bar{\rho}$, and the background \cref{background_H}, we can find the wave-equation for the second-order tensor mode,
\begin{equation}
\label{eom_tensor}
    h^{i(2)''}{_{j}} + 2\mathcal{H}h^{i(2)'}{_{j}} - \nabla^{2}h^{i(2)}{_{j}} = -4P^{li}_{jm}S^{m}{_{l}}, 
\end{equation}
where the source term is given in \cref{source_term}. The second-order tensor perturbation \( h^{(2)}_{ij} \) is sourced by quadratic combinations of first-order scalar modes such as \( \phi^{(1)} \), \( \psi^{(1)} \), and their spatial derivatives. We decided to find this generic form of the source term \cref{source_term_generic} as this can be simplified to further understand the behavior in different gauges by imposing the gauge condition.

In the source term for SIGWs, only quadratic combinations of first-order scalar perturbations appear, while purely second-order scalar perturbations are absent. This is not an ad hoc assumption but a direct consequence of the structure of the Einstein equations at second order. The key reason lies in how the second-order tensor perturbations are sourced: they arise from the non-linear coupling of first-order scalar modes through products of perturbations that project onto the transverse-traceless (TT) part of the spatial Einstein equations.

Crucially, scalar perturbations—whether first or second order—affect only the scalar sector in the scalar–vector–tensor (SVT) decomposition. When projecting the Einstein equations onto the transverse-traceless (TT) part, only specific quadratic combinations of first-order scalars contribute, as purely second-order scalars do not source tensor modes due to their spin-0 nature\footnote{The SVT decomposition classifies perturbations by their behavior under spatial rotations: scalars (spin-0), vectors (spin-1), and transverse-traceless tensors (spin-2). See \cite{Carbone:2004iv,Matarrese:1997ay}.}. This explains why the source term in \cref{eom_tensor} includes only first-order combinations.

This structure is a robust feature of second-order cosmological perturbation theory and has been systematically demonstrated in several works, including \cite{Ananda:2006af, Baumann:2007zm}. Therefore, the form of \cref{eom_tensor} is not the result of an additional assumption, but follows from the TT projection of Einstein’s equations and the properties of metric perturbations under SVT decomposition.

It is convenient to solve the \cref{eom_tensor} in Fourier space in which the source term will become a convolution of the first-order scalar perturbation at different wave-numbers. The second-order tensor mode can be written as
\begin{equation}
h^{(2)}_{ij}(\mathbf{x},\tau)=\sum_{\lambda=+,\times}\dfrac{1}{(2\pi)^{3/2}}\int d^3\mathbf{k}\; e^{i\mathbf{k}\cdot\mathbf{x}} h^{(2)}_{\lambda}(\mathbf{k},\tau)e^\lambda_{ij}(\mathbf{k})\;,
    \label{eq: fourier transform of GW}
\end{equation}
where $\lambda=+,\times$ denotes the two GW polarizations and $e^\lambda_{ij}(\mathbf{k})$ are the polarization tensors. The E.o.M for scalar-induced tensor modes can be written in the frequency domain as 
\begin{equation}
\label{wave_equation_second_order_tensor_mode}
    h^{(2)''}(\mathbf{k},\tau) + 2\mathcal{H}h^{(2)'}(\mathbf{k},\tau) + k^{2}h^{(2)}(\mathbf{k},\tau) = S_{\lambda}(\mathbf{k},\tau), 
\end{equation}
where $S_{\lambda}(\Tilde{\tau},\mathbf{k})$ includes the Fourier transformation of the source \cref{source_term_generic} and depends on the evolution of the scalar modes. Any scalar mode can be split into a transfer function $T (k\tau)$, which includes the time evolution of the modes after they re-enter the horizon, and the initial condition $\zeta_\mathbf{k}$, determined by inflation. For example, the scalar potential $\firstphi(\mathbf{k},\tau)$ can be split as 
\begin{equation}
\label{split_scalar_mode}
    \firstphi (\mathbf{k},\tau) = \frac{3(1+\omega)}{5+3\omega}T_{\phi}(k\tau)\zeta_{\mathbf{k}} \overset{\omega \rightarrow 1/3}{=} \frac{2}{3}T_{\phi}(k\tau)\zeta_{\mathbf{k}}. 
\end{equation}
The primordial curvature fluctuations $\zeta_{\mathbf{k}}$ are characterised by the following power spectrum $\mathcal{P}_\zeta(k)$ as 
\begin{equation}
\label{fluctuation_power_spectrum}
    \langle \zeta_{\mathbf{k}}\zeta_{\mathbf{q}} \rangle = \delta^{(3)}(\mathbf{k} + \mathbf{q})\mathcal{P}_{\zeta}(k) = \dfrac{2\pi^2}{k^3}\delta^{(3)}(\mathbf{k} + \mathbf{q})\Delta^2_\zeta(k)\ , 
\end{equation}
where $\Delta^2_\zeta(k)$ is the dimensionless power spectrum. The source term can be written as
\begin{align*}
\label{fourier_sourceterm}
    S_{\lambda}(\mathbf{k},\tau) & = -4 \mathbf{e}^{l}{_{m}}(\mathbf{k})S_{l}^{m}(\mathbf{k}, \tau)\\
    & =  4\int \frac{d^3\mathbf{q}}{(2\pi)^{3/2}}Q_{\lambda}(\mathbf{k}, \mathbf{q})f(|\mathbf{k} - \mathbf{q}|,q,\tau)\zeta_{\mathbf{q}}\zeta_{\mathbf{k} - \mathbf{q}},  \numberthis
\end{align*}
where we define the polarization tensors, $Q_{\lambda}(\mathbf{k}, \mathbf{q}) \equiv \mathbf{e}^{l}_{\lambda m}(\mathbf{k})\mathbf{q}{_{l}}\mathbf{q}^{m}$ and $f(|\mathbf{k} - \mathbf{q}|,q,\tau)$ is the source function, which is found using the transfer functions of the scalar modes. In this review, we consider the production of SIGW to be when the scalar modes re-enter the causal horizon during the radiation-dominated (RD) era. Therefore, we will be performing the remaining calculations in the RD epoch, where the scale-factor behaves as $a \propto \tau$, resulting in $\mathcal{H} \propto \tau^{-1}$ and $w=1/3$. During the radiation epoch the dimensionless power spectrum for the tensor modes is defined by 
\begin{equation}
   \langle h_{\lambda,\ \mathbf{k}}h_{\lambda',\ \mathbf{q}} \rangle = \delta_{\lambda\lambda'}\delta^{(3)}(\mathbf{k} + \mathbf{q})\dfrac{2\pi^2}{k^3}\Delta^2_h(k,\tau)\ , 
\end{equation}
where $\Delta^2_h(k,\tau)$ is  
\begin{equation}
\label{power_spectrum_spherical_coordinates}
    \Delta^2_{h}(k,\tau) = 8 \int^{\infty}{_{0}} dv \int^{1+v}_{|1-v|} du \left(\frac{4v^2 - \left(1-u^2+v^2\right)^2}{4uv}\right)^{2}I^{2}(v,u,x)\Delta^2_{\zeta}(ku)\Delta^2_{\zeta}(kv)\ .
\end{equation}
Notice the change to spherical momentum coordinates where we use dimensionless variables $u=\frac{|\mathbf{k}-\mathbf{q}|}{k}$ and $v=\frac{\mathbf{q}}{k}$. The function, $I (v,u,x)$ is known as the \textit{kernel} and is found in relation to the source-term \cite{Ananda:2006af,Baumann:2007zm,Kohri:2018awv}, 
\begin{equation}
\label{kernel_radiation}
    I(v,u,x) = \int^{x}_{x_i} d\tilde{x}\frac{\tilde{x}}{x}kG_{\mathbf{k}}(\tau, \tilde{\tau}) f(v,u,\tilde{x}),\
\end{equation}
where the Green's function $G_{\mathbf{k}}$ in radiation era is \cite{Kohri:2018awv} 
\begin{equation}
\label{greens_function}
    k G_\mathbf{k}(x, \Tilde{x}) = \sin(x-\Tilde{x})\ , 
\end{equation}
where $x = k\tau$. Here, we adopt the Green's function that solves the tensor wave equation in a radiation-dominated background with causal boundary conditions, such that it vanishes for $\tau < \tilde{\tau}$ \cite{Kohri:2018awv, Baumann:2007zm}. The factor of $k$ in the sine function argument arises from the structure of the Fourier-space solution in conformal time.
The fraction of GW energy density per logarithmic wavelength is defined by
\begin{equation}
\label{spectral_energy_density}
\Omega_{\rm GW}(k,\tau) \equiv \frac{\rho_{\rm GW}(k,\tau)}{\rho_{c}(\tau)} = \frac{1}{48}\left(\frac{k}{a(\tau)H(\tau)}\right)^{2}\sum_{\lambda=+,\times}\overline{{\Delta^2_{h,\lambda}(k,\tau)}}\ ,
\end{equation}
where $\rho_c(\tau)$ is the critical density and he averaged energy density of GWs is expressed as
\begin{equation}
\rho_{GW}(\tau) = \frac{1}{16a^{2}(\tau)\kappa^{2}}\overline{\langle(\nabla h_{ij})^{2}\rangle}. 
\end{equation}
The overline above the power spectrum is the oscillation average, derived from the average of the kernel. While the second-order tensor perturbations can be constructed to be gauge-invariant under certain conditions, the derived observable—the spectral energy density of gravitational waves, $\Omega_{GW}$—can still exhibit gauge dependence. This is particularly evident in gauges that are not well-behaved on small scales, such as the comoving slicing gauge, where the induced GW spectrum can differ from that in the Newton gauge. Therefore, care must be taken in choosing an appropriate gauge when interpreting the physical implications of SIGWs \cite{Domenech:2020xin,Inomata:2019yww,Yuan:2025seu,Lu:2020diy}.

\section{\label{sec:gauge_first_order}Gauge-dependence of first-order metric perturbations}

We now review how quantities in the metric change under coordinate transformation, also known as gauge-transformation. This section clarifies how different choices of time and space coordinates affect the scalar perturbations and introduces gauge-invariant combinations. There are two approaches in studying gauge-transformation, the \textit{active} and \textit{passive} approach. We follow the approach taken by \cite{Mukhanov:1990me,2001PhDT.......328M, Malik:2008im}, which calculates the change in perturbations under coordinate change at a fixed physical point in spacetime. Later we look at the construction of the gauge-invariant variables.

\subsection{\label{subsec:coordinate_transformation}First-order coordinate transformation of cosmological metric perturbations}

The change in coordinate system for the time slicing and spatial threading is 
\begin{align}
    \tilde{\tau} & = \tau + \xi^{0}, \\
    \tilde{x}^{i} & = x^{i} +  \partial_{i}\xi,
\end{align}
where fixing $\xi^{0}$ determines the time-slicing and the spatial coordinate by $\xi$. With these transformations the perturbed line element can be written in terms of the transformed coordinate system. The scalar modes transform as
\begin{align}
\label{phi_transformation}
    \tilde{\phi}^{(1)} & = \firstphi - \mathcal{H}\xi^{0} - (\xi^{0})^{'}, \\ 
\label{psi_transformation}
    \tilde{\psi}^{(1)} & = \firstpsi + \mathcal{H}\xi^{0}, \\ 
\label{w_transformation}
    \tilde{w}^{(1)} & = \firstw + \xi^{0} - \xi^{'}, \\ 
\label{h_transformation}
    \tilde{h}^{(1)} & = \firsth - \xi.
\end{align}
As discussed earlier in \cref{sec:perturbation_theory}, vector and tensor perturbations at first order—\( w^{(1)}_i \), \( h^{(1)}_i \), and \( h^{(1)}_{ij} \)—are neglected due to their minimal contribution to second-order tensor modes.

\subsection{\label{subsec:gauge_invariant_variables}Gauge-invariant combination of first-order scalar perturbations}

In cosmological perturbation theory, many quantities—such as scalar perturbations—are inherently gauge-dependent, meaning their values can change under coordinate transformations. The gauge-invariant approach addresses this by constructing combinations of perturbation variables that remain unchanged under such transformations. These combinations are called gauge-invariant variables and are designed to coincide with the values of the perturbations in a specific gauge. This procedure involves fixing the coordinate freedom using transformation rules such as those shown in \cref{phi_transformation,psi_transformation,w_transformation,h_transformation}.

It is important to distinguish between gauge-invariant variables and physical observables. Gauge-invariant variables are tools within the theoretical framework that help consistently describe perturbations across different gauges, but they are not necessarily directly measurable. In contrast, observables—such as temperature anisotropies in the cosmic microwave background or gravitational wave signals—are quantities that are invariant under coordinate transformations and correspond to physical measurements. As emphasized in \cite{Inomata:2019yww,DeLuca:2019ufz}, not every gauge-invariant quantity is an observable, though all observables must be gauge-invariant.

Some quantities, like the first-order tensor perturbations, are already gauge-independent—they remain the same in all coordinate systems and are directly related to physical observables like gravitational waves. For gauge-dependent quantities, different gauge choices can lead to different gauge-invariant variables, and the governing equations can be reformulated accordingly. The choice of gauge often depends on the problem being studied and the mathematical convenience it offers \cite{Clifton:2020oqx}.

In the following section, we review gauge choices that have been commonly used in the study of scalar-induced gravitational waves (SIGWs) and discuss their implications for theoretical consistency and observational relevance.

\subsubsection{\label{synchronous}Synchronous gauge}

The synchronous gauge is widely used in numerical computations like those performed in Boltzmann codes (e.g. CLASS \cite{lesgourgues2011cosmiclinearanisotropysolving}). This gauge also known as time-orthogonal gauge and \textit{TT gauge}, which is the typical choice of gauge for gravitational wave measurement and used in numerical studies. It is defined by
\begin{equation}
    g_{00}^{(1)} = g_{0i}^{(1)} = 0, 
\end{equation}
which allows us to set $\firstphi = \hat{w}{_{i}}^{(1)} = 0$. This provides a simplification in the dynamical equations, as the cosmic time coordinates of FRW background coincide with the proper time of a observer at a fixed spatial coordinate, i.e. $d\eta = ad\tau$. One disadvantage with this gauge is that it is not possible to formulate a clear gauge-invariant variable to due residual gauge freedom. The gauge-transformation \cref{phi_transformation,w_transformation} give the following
\begin{align}
    \xi^{0}_{syn} &= -\frac{1}{a}\left[\int a\firstphi d\tau - \mathcal{C}_{1}(x)\right]\\
    \xi_{syn} &= \int (\xi^{0}_{syn} - \firstw)d\tau + \mathcal{\hat{C}}_{1}(x)
\end{align}
which shows that the time-slicing is not determined unambiguously and there remains two arbitrary scalar functions of the spatial coordinate $\mathcal{C}_{1}(x)$ and $\mathcal{\hat{C}}_{1}(x)$. Hence, there is no true gauge-invariant variable. 

\subsubsection{\label{poisson}Poisson gauge}

This gauge is defined by three conditions removing certain degrees of freedom; (i) vanishing shear, (ii) transverse condition on vector modes and (iii) transverse and traceless condition on tensor modes. Hence 
\begin{equation}
    \nabla \cdot w = \nabla \cdot h = 0.
\end{equation}
This gauge is also known as zero-shear gauge. Using this gauge, Bardeen \cite{Bardeen:1980kt} first proposed quantities that invariant under transformation. By studying the transformation equations, the corresponding gauge-invariant variables are known as the Bardeen potentials 
\begin{align}
\label{poisson_variable_1}
    \Phi & = \firstphi - \mathcal{H}\sigma - \sigma',\\
\label{poisson_variable_2}
    \Psi & = \firstpsi + \mathcal{H}\sigma,
\end{align}
where $\firstsigma \equiv h^{(1)'} - \firstw$ is the shear potential. In the framework of general perturbation theory, first-order vector and tensor perturbations remain present. However, by imposing the constraints $w_{i} = h_{ij} = 0$, one recovers the Conformal Newtonian gauge. The Conformal Newtonian gauge is typically employed when vector and tensor perturbations are negligible, whereas the Poisson gauge is preferred in scenarios where vector perturbations play a significant role. In our analysis, since we focus on second-order tensor perturbations, we have already disregarded first-order vector and tensor contributions, leading to the same conditions as those in the Conformal Newtonian gauge.

The Poisson gauge is preferred due to its convenience in calculations on small scales. In the absence of anisotropic stress the scalar potentials are equal, i.e. $\Phi = \Psi$, and the governing field equations take a form close to the Newtonian ones.

\subsubsection{\label{spatially_flat}Uniform curvature gauge}

The uniform curvature gauge, also known as the spatially flat gauge, requires the spatial curvature to be unperturbed and homogeneous. This is accomplished by setting
\begin{equation}
    \firstpsi = \firsth = 0,
\end{equation} 
and the corresponding gauge-invariant variables are
\begin{align}
\label{gauge_invariant_phi_variable_uniform_curvature}
    \tilde{\phi}^{(1)} & = \firstphi + \firstpsi - \left(-\frac{\firstpsi}{\mathcal{H}}\right)',\\
\label{gauge_invariant_w_variable_uniform_curvature}
    \tilde{w}^{(1)} & = \firstw -\frac{\firstpsi}{\mathcal{H}} - h^{(1)'}.
\end{align}
This gauge is more convenient when studying observables during the inflationary period \cite{Clifton:2020oqx}. It is beneficial for analyzing scalar field dynamics and matter perturbations while eliminating the influence of metric fluctuations. It was also mentioned in \cite{Domenech:2020xin}, to being one of the ``well-behaved" gauges on small scales for studying SIGWs. 

\section{\label{sec:source_term_gauges}Source function of SIGWs in different gauges}

Here we identify the explicit form of the source term for second-order gravitational waves that arises from the quadratic coupling of first-order scalar modes. This quantity encapsulates the physical mechanism by which density perturbations generate gravitational radiation during horizon re-entry, and its structure is central to SIGW phenomenology. As mentioned in \cref{subsec:second_order_eqs} source term is found in Fourier space as
\begin{align*}
    S_{\lambda}(\mathbf{k},\tau) & = -4 \mathbf{e}^{l}{_{m}}(\mathbf{k})S_{l}^{m}(\mathbf{k}, \tau)\\
    & =  4\int \frac{d^3\mathbf{q}}{(2\pi)^{3/2}}Q_{\lambda}(\mathbf{k}, \mathbf{q})f(|\mathbf{k} - \mathbf{q}|,q,\tau)\zeta_{\mathbf{q}}\zeta_{\mathbf{k} - \mathbf{q}},  \numberthis
\end{align*}
which is a convolution of \cref{source_term_generic} with primordial fluctuations. We analyze the scalar-scalar source term of SIGWs in the gauges mentioned above using two approaches: (i) deriving the source term in the synchronous and Poisson gauges via their gauge conditions, and (ii) employing gauge-invariant variables to study the source term in the uniform curvature gauge. In all derivations, we retain only first-order scalar perturbations \( \phi^{(1)} \), \( \psi^{(1)} \), etc., as sources for the second-order tensor perturbation \( h^{(2)}_{ij} \).  

\subsection{\label{subsec:source_sync}Scalar-scalar source in synchronous gauge}

Using the definition for the shear potential, $\firstsigma$, and the constraint $\firstphi = \firstw = 0$, \cref{source_term_generic} simplifies to, 
\begin{align*}
\label{sourceterm_sync_gr_simplified}
    S^{i}{_{j,S}}  = &  -\dfirstpsi\partial^{i}\partial_{j}\firstsigma + 2\partial^{i}\dfirstpsi\partial_{j}\firstsigma + 2\partial_{j}\dfirstpsi\partial^{i}\firstsigma + \firstpsi\partial^{i}\partial_{j}\firstpsi \\
    & - \frac{1}{\mathcal{H}^{2}}\partial^{i}\dfirstpsi\partial_{j}\dfirstpsi + \partial^{i}\partial_{j}\firstsigma\nabla^{2}\firstsigma - \partial^{i}\partial_{k}\firstsigma\partial^{k}\partial_{j}\firstsigma. \numberthis
\end{align*}
In Fourier space, the source function becomes
\begin{align*}
\label{f_source_term_sync}
    f_{S}(|\mathbf{k} - \mathbf{q}|,q,\tau) = & \frac{4}{9}\left[T_{\psi}(q\tau)T_{\psi}(|\mathbf{k} - \mathbf{q}|\tau) - 2T'_{\psi}(q\tau)T_{\sigma}(|\mathbf{k} - \mathbf{q}|\tau)  \right. \\
    & \left. - T_{\sigma}(q\tau)T'_{\psi}(|\mathbf{k} - \mathbf{q}|\tau) + \tau^{2}T'_{\psi}(q\tau)T'_{\psi}(|\mathbf{k} - \mathbf{q}|\tau) \right. \\
    & \left. - (\mathbf{k} - \mathbf{q})T_{\sigma}(q\tau)T_{\sigma}(|\mathbf{k} - \mathbf{q}|\tau) - (\mathbf{k}\cdot\mathbf{q} - q^{2})T_{\sigma}(q\tau)T_{\sigma}(|\mathbf{k} - \mathbf{q}|\tau)\right ]. \numberthis
\end{align*}

\subsection{\label{subsec:source_poisson}Scalar-scalar source in Poisson gauge}

By using the gauge-condition $\firstw = \firsth = 0$ and in the absence of anisotropic stress where $\firstphi = \firstpsi$, we can simplify \cref{source_term_generic} in the Poisson gauge 
\begin{align*}
\label{sourceterm_poisson_gr_simplified}
    S^{i}{_{j,P}} & = 4\firstphi\partial^{i}\partial_{j}\firstphi + \partial^{i}\firstphi\partial_{j}\firstphi  \\
    & - \tau^{2}\left(\partial^{i}\dfirstphi\partial_{j}\dfirstphi + \tau^{-1}\partial^{i}\dfirstphi\partial_{j}\firstphi + \tau^{-1}\partial^{i}\firstphi\partial_{j}\dfirstphi\right). \numberthis
\end{align*}
Therefore the source function in Fourier space is
\begin{align*}
\label{f_source_term_poisson}
    f_{P}(|\mathbf{k} - \mathbf{q}|,q,\tau) = & \frac{4}{9}\left[3 T_{\phi}(q\tau)T_{\phi}(|\mathbf{k} - \mathbf{q}|\tau) + \tau^{2}T'_{\phi}(q\tau)T'_{\phi}(|\mathbf{k} - \mathbf{q}|\tau) \right. \\
    & \left. + \tau \left\{T_{\phi}(q\tau)T'_{\phi}(|\mathbf{k} - \mathbf{q}|\tau) + T_{\phi}(q\tau)T'_{\phi}(|\mathbf{k} - \mathbf{q}|\tau)\right\}\right]. \numberthis
\end{align*}

\subsection{\label{subsec:source_uniform}Scalar-scalar source in uniform curvature gauge}

Next we will study the uniform curvature gauge where $\firstpsi = \firsth = 0$. We can reduce \cref{source_term_generic} to 
\begin{align*}
\label{sourceterm_spatially_flat_simplified}
    S^{i}{_{j,U}} &= 2\firstphi\partial^{i}\partial_{j}\firstphi + 4\tau^{-1}\firstphi\partial^{i}\partial_{j}\firstw + \dfirstphi\partial^{i}\partial_{j}\firstw + 2\firstphi\partial^{i}\partial_{j}\dfirstw \\
    & + \nabla^{2}\firstw\partial^{i}\partial_{j}\firstw - \partial^{i}\partial_{k}\firstw\partial^{k}\partial_{j}\firstw. \numberthis
\end{align*}
The standard way would be to find the solution for $\firstphi$ and $\firstw$ and then later solve the E.o.M for SIGW. On the other hand, we have the gauge-invariant variables in the uniform curvature gauge, \cref{gauge_invariant_phi_variable_uniform_curvature} and \cref{gauge_invariant_w_variable_uniform_curvature}. We can therefore use this to our advantage. First we can make the same assumption of $\firstpsi=\firstphi$ as done in the Poisson gauge and re-write the gauge-invariant variables as
\begin{align}
\label{gauge_invariant_phi_uniform_curvature_simplified}
    \tilde{\phi}^{(1)} & = 2\firstphi - \left(-\frac{\firstphi}{\mathcal{H}}\right)',\\
    \tilde{w}^{(1)} & = - \frac{\firstphi}{\mathcal{H}}.
 \label{gauge_invariant_w_uniform_curvature_simplified}
\end{align}
As done in the Poisson gauge, the source term can be written in Fourier space in terms of the gauge-invariant variables and the transfer functions for the scalar potential, $\firstphi$, 
\begin{align*}
\label{f_source_term_uniform}
    f_{U}(|\mathbf{k} - \mathbf{q}|,q,\tau) = & \frac{4}{9}\left[ 4\tau T_{\phi}(q\tau)T'_{\phi}(|\mathbf{k} - \mathbf{q}|\tau) - \tau^{2}T''_{\phi}(q\tau)T_{\phi}(|\mathbf{k} - \mathbf{q}|\tau) \right. \\
    & \left. - 8\tau T'_{\phi}(q\tau)T_{\phi}(|\mathbf{k} - \mathbf{q}|\tau)  - \tau^{2}(\mathbf{k} - \mathbf{q})T_{\phi}(q\tau)T_{\phi}(|\mathbf{k} - \mathbf{q}|\tau) \right. \\
    & \left. - \tau^{2}(\mathbf{k}\cdot\mathbf{q} - q^{2})T_{\phi}(q\tau)T_{\phi}(|\mathbf{k} - \mathbf{q}|\tau)\right]. \numberthis
\end{align*}
This means that we will use the solution for the scalar-potential in the Poisson gauge to evaluate the source function in the uniform curvature gauge.  

\section{\label{sec:gauges_kernel}Kernel in various gauges}

In this section we derive the analytical expressions of the source function and later perform numerical integration of the kernel in the various gauges. As seen in \cref{sec:source_term_gauges}, the source function in the synchronous gauge, \cref{f_source_term_sync}, is found in terms of the transfer functions for the scalar modes $T_{\psi}$ and $T_{\sigma}$. We study the behavior in synchronous gauge, following the method outlined in \cite{DeLuca:2019ufz}, by deriving the linear solutions for the scalar modes and subsequently simplify the source term. On the other hand, the source function in the Poisson gauge, \cref{f_source_term_poisson}, and uniform gauge, \cref{f_source_term_uniform}, are written in terms of the transfer function for the scalar potential $\firstphi$. We can repeat the same analysis of the first-order scalar potential in the Poisson gauge. Using the transfer function, $T_{\phi}$, we are able to compute both in the Poisson and uniform gauge. For the numerical analysis, we use the first-order transfer functions for the scalar perturbations, to derive the analytical expression of the source function in all three gauges. We present the source functions in Appendix \ref{subsec:analytical_kernel} in terms of the transfer functions, where the transfer functions \( T_\psi(k, \tau) \), \( T_\phi(k, \tau) \), and \( T_\sigma(k, \tau) \) describe the time evolution of first-order scalar perturbations in the respective gauges.

\subsection{\label{subsec:solution_sync}Linear solutions in synchronous gauge}

We apply the definition $\firstphi = \firstw = 0$ to the first-order equations \cref{00_component_evolution}--\cref{ij_component_traceless} and using the definition for shear, $\firstsigma$, the equations of motion for $\firstpsi$ and $\firstsigma$ is obtained
\begin{align}
    2\mathcal{H}\firstsigma + \dfirstsigma + \firstpsi &= 0, \\
    6\ddfirstpsi + 2\mathcal{H}\left(9\dfirstpsi - 4\nabla^{2}\firstsigma\right) - 3\nabla^{2}\dfirstsigma - 5\nabla^{2}\firstpsi & = 0. 
\end{align}
The solutions are given as 
\begin{equation}
\label{first_order_solution_synchronous_psi}
   T_{\psi}(k,\tau) = 9\left(\frac{\sin\left(\frac{k\tau}{\sqrt{3}}\right) - k\tau}{(k\tau)^{2}}\right),
\end{equation}
\begin{equation}
\label{first_order_solution_synchronous_sigma}
    T_{\sigma}(k,\tau) =  9\left(\frac{1-\cos\left(\frac{k\tau}{\sqrt{3}}\right)}{(k\tau)^{2}}\right).\\
\end{equation}

\subsection{\label{subsec:solution_poisson}Linear solutions in Poisson gauge}

As mentioned in the previous sections, this gauge is defined by $ \firstw = \firsth = 0$. In the first place we see that the linear-order trace-free part of the spatial Einstein field equation, \cref{ij_component_traceless} provides us with the following constraint equation
\begin{equation}
\label{constraint_equation_anisotropic_stress}
	\left(\partial^{i}\partial_{j} - \frac{1}{3}\nabla^{2}\delta^{i}{_{j}}\right)\left(\firstphi - \firstpsi\right) = \kappa^{2}\pi^{(1)i}{_{j}}\ ,
\end{equation}
The source anisotropic stress comes from free-streaming neutrinos which is minimal and therefore we can neglect $\pi^{(1)i}{_{j}}$ and set $\firstpsi = \firstphi$. The linear-order solution for the scalar potential is found using the first-order evolution equations, \cref{00_component_evolution} and \cref{ij_component_trace}
\begin{equation}
\label{simplified_first_order_evolution_equation}
    \phi^{(1)''} + 3\mathcal{H}\left(1 + c_{s}^{2}\right)\phi^{(1)'} + \left(2\mathcal{H}' + \left(1 + 3c_{s}^{2}\right)\mathcal{H}^{2} + c_{s}^{2}k^{2}\right)\firstphi = 0\ ,
\end{equation}
The general solution in RD era is found as Bessel function,
\begin{equation}
\label{approximate_solution_phi_radiation}
    T_{\phi}(k,\tau) = \frac{9}{(k\tau)^{2}}\left[\frac{\sqrt{3}}{k\tau}\sin\left(\frac{k\tau}{\sqrt{3}}\right) - \cos\left(\frac{k\tau}{\sqrt{3}}\right)\right].
\end{equation} 

\subsection{\label{subsec:numerical_analysis}Numerical integration of the kernel}

\begin{figure}[h!]
    \centering  
    \includegraphics[width=1\textwidth]{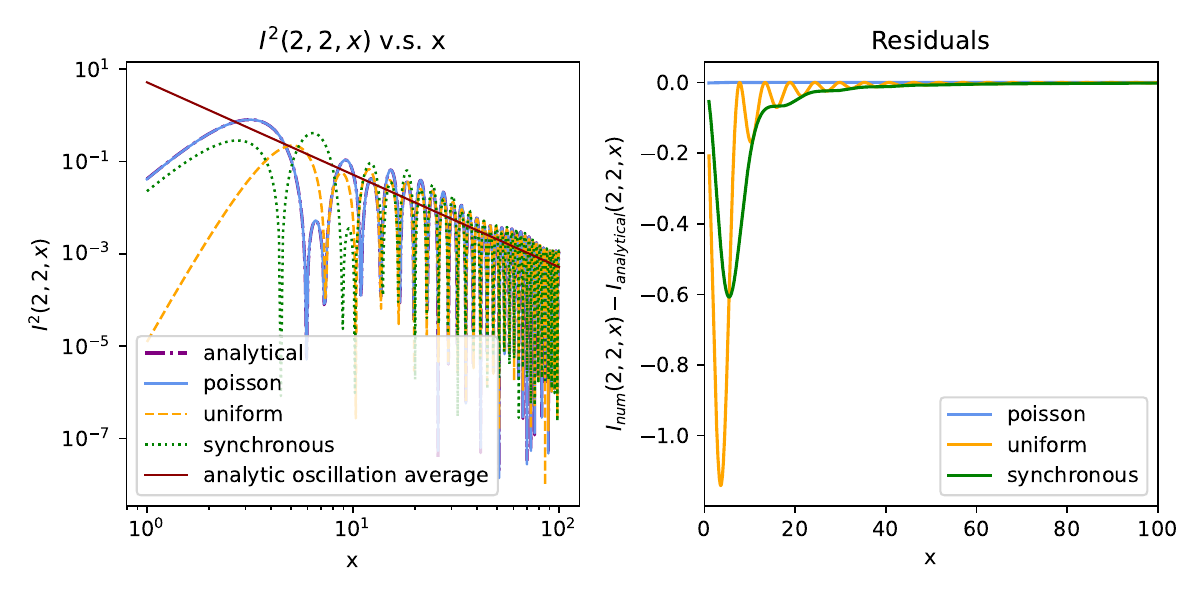}  
    \caption{The kernel function $I^{2}(v=1,u=1,x)$ as a function of $x=k\tau$. In the left panel we show the squared analytical kernel (purple line) taken from Eq. (25), and it's oscillation average (red line) taken from Eq.(26) of \cite{Kohri:2018awv} compared to numerical integration of the kernel in the synchronous (green line), Poisson (blue line) and uniform curvature (orange line) gauges. In the right panel, we plot the residual between the numerically integrated kernel in three gauges and the analytical integration Eq. (25) from \cite{Kohri:2018awv}.}  
    \label{fig:numerical_kernel}  
\end{figure}

We numerically evaluate the kernel, \cref{kernel_radiation}, in the three different gauges using the source functions \cref{analytical_source_sync}, \cref{analytical_source_poisson} and \cref{analytical_source_uniform}. By fixing a set of arbitrary values for $u = 1$ and $v = 1$, the numerical integration is presented in 
Fig.~\ref{fig:numerical_kernel}. The parameters $u$ and $v$ determine the time of horizon entry of the source modes, with higher values corresponding to modes that enter much later. In the radiation-dominated era, the transfer function scales approximately as $q\tau$, and hence becomes suppressed at late times. As a result, configurations with large $u$ and $v$ contribute negligibly to the kernel. To capture the most significant contribution to the gravitational wave signal, we focus primarily on small values of $u$ and $v$, where the sources are most active. We compare the numerically integrated kernel in all three gauges with the analytical squared kernel from Eq. (25) and it's oscillation-averaged from Eq. (26) of \cite{Kohri:2018awv}, derived in the conformal Newtonian gauge. This is shown in the left panel, where the numerical kernels are compared to the analytical squared kernel. In the right panel, we show the residuals between the numerical squared kernels and the analytical squared kernel from Eq. (25) of \cite{Kohri:2018awv}. 

We observe that the numerical results from the different gauge choices closely follow the analytical solution. The kernel values for the Poisson, uniform curvature, and synchronous gauges exhibit minimal variation, indicating that any residual gauge effects are small. The oscillatory structure seen in the analytical oscillation average aligns well with the numerically integrated kernels. From the right panel, we see that the residuals remain small across all gauges, supporting the conclusion that the choice of gauge does not significantly affect the final gravitational wave kernel. In the region \( x \sim 50 \) to \( x \sim 100 \), a modulation in the oscillation amplitude of the numerical kernel becomes apparent, particularly in the synchronous and uniform curvature gauges. This modulation occurs on a scale larger than the individual oscillation period. We interpret this effect as a numerical artefact rather than a physical feature. The modulation is likely related to the numerical integration scheme, which may imperfectly capture the rapidly oscillating integrand at late times.

\begin{figure}[h!]
    \centering  
    \includegraphics[width=1\textwidth]{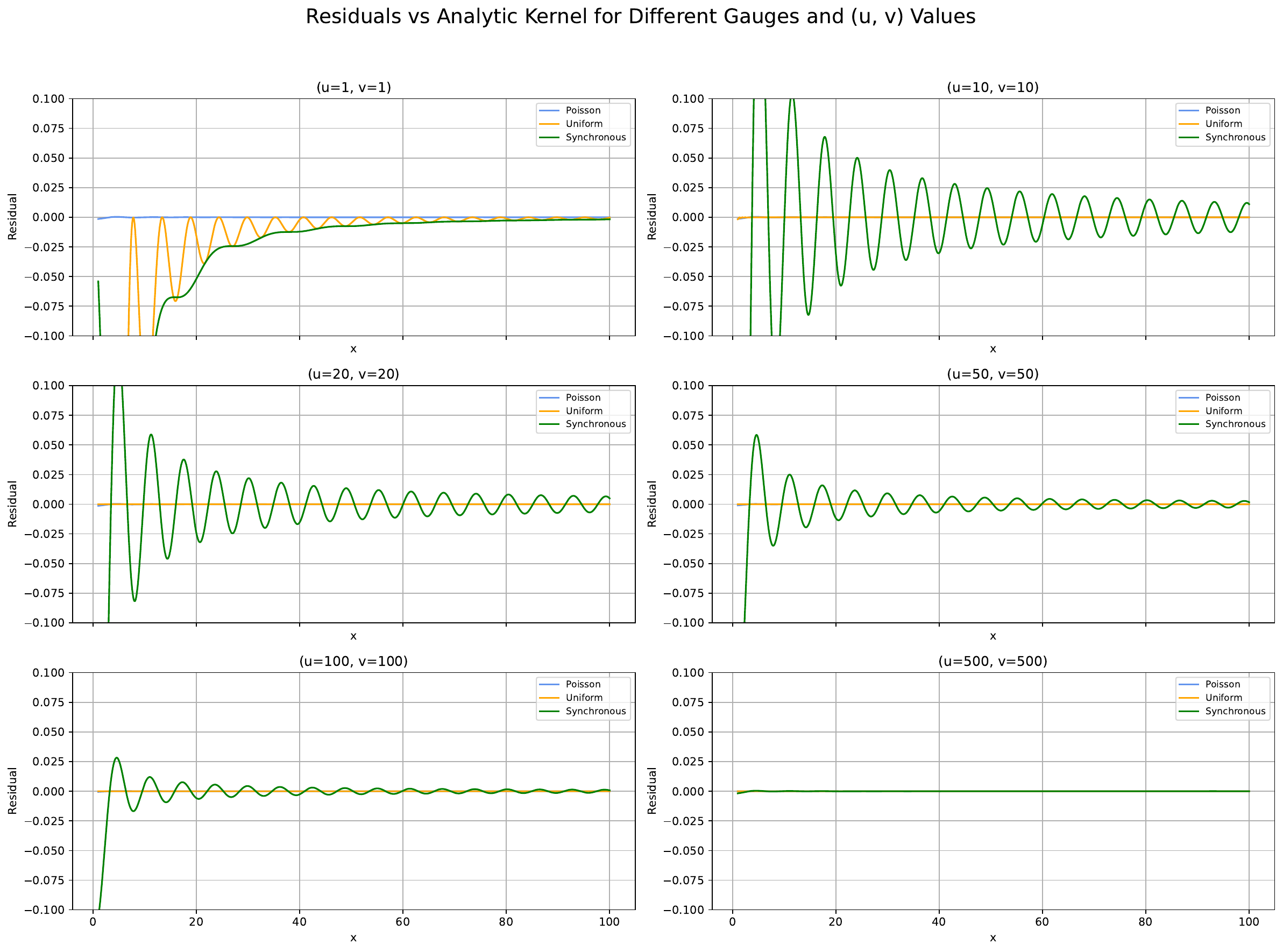}  
    \caption{Residuals between the numerically integrated kernels and the analytical expression from Eq.~(25) of \cite{Kohri:2018awv}, plotted for different values of the parameters \( u \) and \( v \). Each subplot corresponds to a different pair of \( (u, v) \), illustrating how the residual structure varies with the choice of source mode configuration.}  
    \label{fig:numerical_residuals}  
\end{figure}

For larger values of \( x \), the synchronous gauge (green curve) appears to lie below the uniform curvature gauge (orange curve), with points of tangency occurring near the oscillation minima of the orange curve. However, this behavior is not systematic and should not be interpreted as a physical lower bound. Upon closer inspection and using higher-resolution plots or narrower vertical limits, the apparent tangency is revealed to be an artifact of visual resolution rather than a genuine feature of the underlying functions. Moreover, the residuals depend on the specific values of the dimensionless variables \( u \) and \( v \). We illustrate this dependence in Fig.~\ref{fig:numerical_residuals}, which shows how the residual structure varies significantly with different choices of $u$ and $v$. We also see that the residual between the numerical and analytical results tends to decrease for larger values of \( u \) and \( v \), as the transfer function becomes increasingly suppressed and the numerical kernel approaches the analytical approximation more closely.

Even if there is a minimal discrepancy, the variation in behavior in all three gauges decreases as $x \gg 1$, when entering deep in the horizon. 

\subsection{Connecting SIGWs to Present-Day Observables}

To assess the observational relevance of scalar-induced gravitational waves (SIGWs), we must relate their spectral energy density at the time of generation during the radiation-dominated (RD) epoch to their present-day energy density spectrum, \( \Omega_{\rm GW,0}(k) \). This connection is established under two primary assumptions:

\begin{enumerate}
    \item The emission of SIGWs occurs after the reheating phase and during the RD epoch, which is when curvature perturbations re-enter the horizon and induce tensor modes.
    \item After their generation, the energy density of the GWs redshifts as radiation, \( \rho_{\rm GW} \propto a^{-4} \), until matter-radiation equality. Post-equality, this radiation dilution continues in the presence of non-negligible changes in the relativistic degrees of freedom, encoded through the functions \( g_* \) and \( g_{*s} \), representing the effective number of relativistic species and entropy degrees of freedom respectively.
\end{enumerate}

Taking these effects into account, the present-day GW spectral density is given by~\cite{Watanabe:2006qe}:
\begin{equation}
    \Omega_{\rm GW,0}(k)h^{2} = \Omega_{r,0}h^{2} \left(\frac{g_{*s}(T_{\rm hc})}{g_{*s0}}\right)^{-4/3} \left(\frac{g_{*}(T_{\rm hc})}{g_{*0}}\right) \Omega_{\rm GW,hc}(k)\,,
\end{equation}
where \( h = H_0/100\,\text{km}\,\text{s}^{-1}\,\text{Mpc}^{-1} \) is the reduced Hubble constant, and the subscript ``hc'' denotes quantities evaluated at the horizon-crossing time during the RD era. For temperatures \( T_{\rm hc} \gtrsim 1\,\mathrm{TeV} \), the Standard Model gives \( g_* = g_{*s} = 106.75 \). The current radiation density parameter is constrained by Planck measurements to be \( \Omega_{r,0} h^2 \approx 4.2 \times 10^{-5} \)~\cite{Planck:2018vyg}.

The only remaining input required to perform the numerical integration of the SIGW spectrum is the primordial curvature power spectrum. Since the SIGW spectrum scales quadratically with the primordial scalar power spectrum, we adopt a log-normal form for the dimensionless spectrum~\cite{Pi:2020otn}:
\begin{equation}
    \label{eq:scalar_power_spectrum_log_normal}
    \Delta^2_{\zeta}(k) = \frac{\mathcal{A}_{\zeta}}{\sqrt{2\pi\sigma^{2}}}\exp\left(-\frac{\ln^{2}({k/k_{*}})}{2\sigma^{2}}\right)\,,
\end{equation}
where \( \mathcal{A}_\zeta \) is the total integrated amplitude, \( \sigma \) determines the width of the peak, and \( k_* \) its position. In our computations, we use \( \mathcal{A}_\zeta = 2.1 \times 10^{-2} \) and \( \sigma = 0.1 \), which ensures a narrow peak structure significantly above the CMB amplitude (\( A_\zeta(k_{\rm CMB}) \sim 2.1 \times 10^{-9} \) at \( k_{\rm CMB} = 0.05\,\mathrm{Mpc}^{-1} \)), while remaining unconstrained at smaller scales relevant for SIGWs. The peak wavenumber \( k_* \) can be chosen based on the detector’s sensitivity using the relation \( f_{*} = k_{*}/(2\pi) \).

\begin{figure}[h!]
    \centering
    \includegraphics[width=1\textwidth]{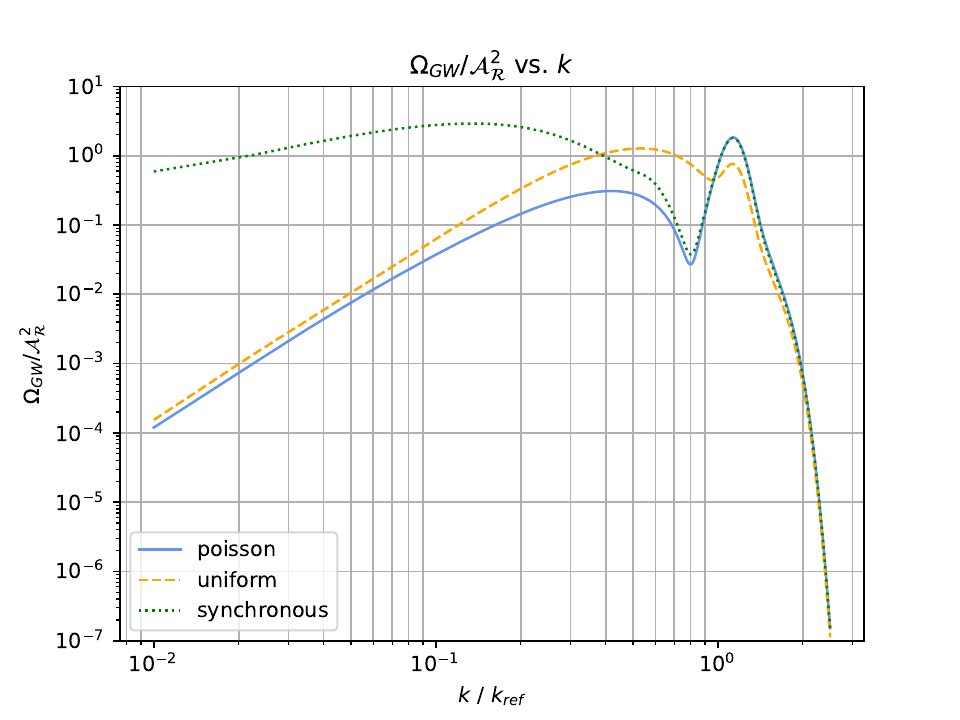}
    \caption{Spectral energy density of scalar-induced gravitational waves (SIGWs), \( \Omega_{\rm GW}(k)/\Omega_r \), as a function of comoving wavenumber normalized to a reference scale \( k/k_{\rm ref} \). Results are shown for three gauge choices: Poisson (blue), synchronous (green), and uniform curvature (orange), computed during the radiation-dominated era.}
    \label{fig:omega_gw_gauges}
\end{figure}

In Fig.~\ref{fig:omega_gw_gauges}, we compare the SIGW energy density spectrum computed in three commonly used gauges—Poisson, synchronous, and uniform curvature.
The three curves show good agreement, particularly in the sub-horizon regime (\( k \gg k_{\rm ref} \)), where the scalar source term becomes negligible and gravitational waves propagate freely. As expected, gauge dependence is most pronounced on super-horizon scales, where the scalar source remains active and its gauge-specific formulation directly affects the evolution of tensor modes. Minor discrepancies at intermediate scales can be attributed to residual gauge effects from numerical integration and the gauge-dependent structure of the source term. These variations diminish at late times, reinforcing the conclusion that the SIGW spectrum is effectively gauge-invariant in the observationally relevant regime. This supports earlier arguments that gauge effects are suppressed once tensor modes decouple from their scalar sources and evolve independently \cite{DeLuca:2019ufz,Domenech:2020xin}.

\section{\label{sec:conclusion}Conclusion}

In this review, we have revisited the issue of gauge-dependence of ``scalar-induced" gravitational waves, which are second-order gravitational waves produced by first-order scalars. We approach this issue by addressing the gauge-invariant approach which has already been discussed in previous studies \cite{Hwang:2017oxa, Tomikawa:2019tvi, Yuan:2019fwv, DeLuca:2019ufz, Domenech:2020xin}. We start by studying the metric perturbation up to second-order without imposing a gauge and retrieve the scalar-scalar source term for the second-order tensor perturbations. Next we revise the gauge-invariant approach by studying the gauge-transformation properties and deriving the gauge-invariant variables for different gauges; Poisson, uniform curvature and synchronous. We adopt two approaches in finding the behavior in the different gauge. For the synchronous gauge, we find the subsequent first-order solutions for the perturbations in that gauge and compute the kernel. For the analysis in Poisson and uniform curvature gauge, we use the gauge invariant method by initially finding the first-order solution in the Poisson gauge and then following similar procedure to \cite{Yuan:2019fwv}, we use the gauge-invariant variables found in \cref{spatially_flat} to then find the kernel in the uniform curvature gauge. Our computation has been done numerically in the radiation-dominant case, with $w=1/3$. 

The results confirm that the kernel gravitational waves remains robust across different gauge choices similar to findings from \cite{Gong:2019mui, Lu:2020diy}. The numerical integration in the Poisson gauge successfully reproduces the expected analytical behavior from \cite{Kohri:2018awv}, validating the computational approach, since the Poisson gauge reduces to the Conformal Newtonian gauge in the absence of vector perturbations. As discussed previously, the Poisson gauge is often preferred for such calculations due to its direct relation to Newtonian-like potentials, while the synchronous gauge can introduce coordinate artifacts due to its residual gauge freedom.

It is important to emphasize that the approximate gauge invariance we observe applies specifically in the radiation-dominated epoch well inside the horizon ($x \gg 1$) when the scalar sources become inactive. As demonstrated in \cite{Domenech:2020xin}, in a dust-dominated universe or during epochs where sources remain active, residual gauge effects can persist and impact observables. Therefore, the gauge-invariant behavior of SIGWs should be interpreted within the appropriate cosmological context.

\bmhead{Acknowledgments}
I would like to thank S. Matarrese, A. Maselli and G. Perna for their insightful comments and discussions. My gratitude also goes to M. Traforetti for her assistance with the computation and numerical integration of the kernels.

\begin{appendices}

\section{Ricci tensor}\label{ricci_tensor}

The Ricci tensor, which is a contraction of the Riemann tensor, can be written in terms of the connection coefficient

\begin{equation}
	R_{\mu\nu} = \partial_{\alpha}\Gamma^{\alpha}_{\mu\nu} - \partial_{\nu}\Gamma^{\alpha}_{\nu\alpha} + \Gamma^{\alpha}_{\sigma\alpha}\Gamma^{\sigma}_{\mu\nu} - \Gamma^{\alpha}_{\sigma\nu}\Gamma^{\sigma}_{\mu\alpha},
\end{equation} 

where the connection coefficients is defined as

\begin{equation}
	\Gamma^{\alpha}_{\beta\gamma} = \frac{1}{2}g^{\alpha\sigma}\left(\frac{\partial g_{\sigma\gamma}}{\partial x^{\beta}} + \frac{\partial g_{\beta\sigma}}{\partial x^{\gamma}} - \frac{\partial g_{\beta\gamma}}{\partial x^{\sigma}}\right).
\end{equation} 

The perturbed components are, 

\begin{align*}
\label{zeroth_order_R_tensor}
    R^{(0)}_{00} & = -3\frac{a''}{a} + 3\left(\frac{a'}{a}\right)^{2}, \numberthis \\
    R^{(0)}_{0i} & = 0, \numberthis \\
    R^{(0)}_{ij} & = \left[\frac{a''}{a} +\left(\frac{a'}{a}\right)^{2}\right]\delta^{i}{_{j}}. \numberthis
\end{align*}

\begin{align*}
\label{first_order_R_tensor}
    R^{(1)}_{00} & = \nabla^{2}\firstphi + 3\frac{a'}{a}\dfirstphi + 3\ddfirstpsi + 3\frac{a'}{a}\dfirstpsi + \frac{a'}{a}\nabla^{2}\firstw + \nabla^{2}\dfirstw, \numberthis \\
    R^{(1)}_{0i} & = 2\frac{a'}{a}\partial_{i}\firstphi + 2\partial_{i}\dfirstpsi + \left(\frac{a''}{a}+\left(\frac{a'}{a}\right)^{2}\right)\partial_{i}\firstw + \frac{1}{2}\partial_{k}D^{k}{_{i}}\firsth, \numberthis \\
    R^{(1)}_{ij} & = \left[-2\left(\frac{a''}{a} +\frac{a'}{a}\right)^{2}\firstphi -\frac{a'}{a}\dfirstphi -2\left(\frac{a''}{a} +\frac{a'}{a}\right)^{2}\firstpsi - 5\frac{a'}{a}\dfirstpsi \right. \\
    & \left. - \ddfirstpsi + \partial^{k}\partial_{k}\firstpsi - \frac{a'}{a}\nabla^{2}\firstw\right]\delta^{i}{_{j}}  + \partial_{i}\partial_{j}\firstpsi - \partial_{i}\partial_{j}\firstphi \\
    & - 2\frac{a'}{a}\partial_{i}\partial_{j}\firstw - \partial_{i}\partial_{j}\dfirstw  + \left(\frac{a''}{a} +\frac{a'}{a}\right)^{2}D_{ij}\firsth + \frac{a'}{a}D_{ij}\dfirsth \\
    & + \frac{1}{2}D_{ij}\ddfirsth + \frac{1}{2}\partial_{i}\partial_{k}D^{k}{_{j}}\firsth + \frac{1}{2}\partial_{j}\partial_{j}D^{k}{_{i}}\firsth - \frac{1}{2}\nabla^{2}D_{ij}\firsth. \numberthis 
\end{align*}

\begin{align*}
\label{second_order_R_tensor_00}
    R^{(2)}_{00} & = \frac{3}{2}\frac{a'}{a}\phi^{(2)'}+\frac{1}{2}\nabla^{2}\phi^{(2)} + \frac{3}{2}\frac{a'}{a}\psi^{(2)'} + \frac{3}{2}\psi^{(2)''}  + \frac{1}{2}\frac{a'}{a}\nabla^{2}w^{(2)} + \frac{1}{2}\nabla^{2}w^{(2)'} \\
    & - 6\frac{a'}{a}\firstphi\dfirstphi - \partial^{k}\firstphi\partial_{k}\firstphi - 3\dfirstphi\dfirstpsi + 2\firstpsi\nabla^{2}\firstphi - \partial_{k}\firstpsi\partial^{k}\firstphi \\
    & + 6\frac{a'}{a}\firstpsi\dfirstpsi + 6\firstpsi\ddfirstpsi + 3(\dfirstpsi)^{2} - \dfirstphi\nabla^{2}\firstw + \frac{a'}{a}\partial_{k}\firstphi\partial^{k}\firstw \\
    & - \frac{a'}{a}\partial_{k}\firstpsi\partial^{k}\firstw + 2\frac{a'}{a}\firstpsi\nabla^{2}\firstw - \partial_{k}\firstpsi\partial^{k}\dfirstw + 2\firstpsi\nabla^{2}\dfirstw  \\
    & + \left(\frac{a''}{a}+\left(\frac{a'}{a}\right)^{2}\right)\partial_{k}\firstw\partial^{k}\firstw + 3\frac{a'}{a}\partial_{k}\dfirstw\partial^{k}\firstw  - \partial_{k}\firstphi\partial_{m}D^{km}\firsth \\
    & - \partial_{k}\partial_{m}\firstphi D^{km}\firsth  - \frac{a'}{a}\partial_{k}\firstw\partial_{m}D^{km}\firsth  - \frac{a'}{a}\partial_{k}\partial_{m}\firstw D^{km}\firsth \\
    & - \partial_{k}\dfirstw\partial_{m}D^{km}\firsth - \partial_{k}\partial_{m}\dfirstw D^{km}\firsth  - \frac{1}{2}\frac{a'}{a}D^{km}\firsth D_{km}\dfirsth  \\
    & + \frac{1}{4}D^{km}\dfirsth D_{km}\dfirsth + \frac{1}{2}D^{km}\firsth D_{km}\ddfirsth, \numberthis
\end{align*}

\begin{align*}
\label{second_order_R_tensor_0i}
    R^{(2)}_{0i} & = \frac{a'}{a}\partial_{i}\phi^{(2)} + \partial_{i}\psi^{(2)'} - \frac{1}{4}\nabla^{2}w_{i}^{(2)} + \frac{1}{2}\left[\frac{a''}{a} +\left(\frac{a'}{a}\right)^{2}\right]w_{i}^{(2)} \\
    & + \frac{1}{2}\left[\frac{a''}{a}  + \left(\frac{a'}{a}\right)^{2}\right]\partial_{i}w^{(2)}  - \frac{1}{2}\nabla^{2}h^{(2)'}{_{i}} + \frac{1}{4}\partial_{k}D^{k}{_{i}}h^{(2)'} \\
    & - 4\frac{a'}{a}\firstphi\partial_{i}\firstphi - 2\dfirstpsi\partial_{i}\firstphi + 4\dfirstpsi\partial_{i}\firstpsi  + 4\firstpsi\partial_{i}\dfirstpsi  \\
    & - 2\left[\frac{a''}{a} +\left(\frac{a'}{a}\right)^{2}\right]\firstphi\partial_{i}\firstw - \partial_{i}\firstphi\nabla^{2}\firstw \\
    & - \partial_{i}\partial_{k}\firstphi\partial^{k}\firstw + \partial^{k}\firstphi\partial_{i}\partial_{k}\firstw - \frac{a'}{a}\dfirstphi\partial_{i}\firstw \\
    & - 5\frac{a'}{a}\dfirstpsi\partial_{i}\firstw - \ddfirstpsi\partial_{i}\firstw - \frac{a'}{a}\partial_{i}\firstw\nabla^{2}\firstw - \partial^{k}\firstw\partial_{i}\partial_{k}\dfirstw \\
    & - \frac{1}{2}\partial_{k}\firstphi D^{k}{_{i}}\dfirsth + \frac{1}{2}\partial_{k}\firstpsi D^{k}{_{i}}\dfirsth + \firstpsi\partial_{k}D^{k}{_{i}}\dfirsth \\
    & + \partial_{k}\dfirstpsi D^{k}{_{i}}\firsth + \dfirstpsi\partial_{k}D^{k}{_{i}}\firsth + \frac{a'}{a}\partial^{k}\firstw D_{ik}\dfirsth \\
    & + \frac{1}{2}\partial^{k}\firstw D_{ik}\ddfirsth  - \frac{1}{2}D^{km}\firsth\partial_{k}D_{mi}\dfirsth + \frac{1}{2}D^{km}\firsth\partial_{i}D_{km}\dfirsth \\
    & - \frac{1}{2}\partial_{k}D^{km}\firsth D_{mi}\dfirsth + \frac{1}{4}\partial_{i}D^{km}\firsth D_{km}\dfirsth, \numberthis
\end{align*}

\begin{align*}
\label{second_order_R_tensor_ij}
    R^{(2)}_{ij} & = \left[-\left(\frac{a''}{a} +\left(\frac{a'}{a}\right)^{2}\right)\phi^{(2)} - \frac{1}{2}\frac{a'}{a}\phi^{(2)'} -\left(\frac{a''}{a} +\left(\frac{a'}{a}\right)^{2}\right)\psi^{(2)}  - \frac{5}{2}\frac{a'}{a}\psi^{(2)'} \right. \\
    & \left. - \frac{1}{2}\psi^{(2)''} +  \frac{1}{2}\nabla^{2}\psi^{(2)} - \frac{1}{2}\frac{a'}{a}\nabla^{2}w^{(2)} + 4\left(\frac{a''}{a} +\left(\frac{a'}{a}\right)^{2}\right)(\firstphi)^{2} \right. \\
    & \left. + 4\frac{a'}{a}\firstphi\dfirstphi + 19\frac{a'}{a}\firstphi\dfirstpsi + 2\frac{a'}{a}\dfirstphi\firstpsi + \dfirstphi\dfirstpsi + 2\firstphi\ddfirstpsi \right. \\
    & \left. 4\left(\frac{a''}{a} +\left(\frac{a'}{a}\right)^{2}\right)\firstphi\firstpsi + \partial_{k}\firstpsi\partial^{k}\firstphi + (\dfirstpsi)^{2} + \partial_{k}\firstpsi\partial^{k}\firstpsi \right. \\
    & \left.  + 2\firstpsi\nabla^{2}\firstpsi + \frac{a'}{a}\partial_{k}\firstphi\partial^{k}\firstw  + 2\frac{a'}{a}\firstphi\nabla^{2}\firstw \right. \\
    & \left. + 3\frac{a'}{a}\partial^{k}\firstpsi\partial_{k}\firstw + \partial_{k}\firstpsi\partial^{k}\dfirstw  + 2\partial_{k}\dfirstpsi\partial^{k}\firstw \right. \\
    & \left. + \dfirstpsi\nabla^{2}\firstw - \partial_{m}\firstpsi\partial_{k}D^{km}\firsth  - \partial_{k}\partial_{m}\firstpsi D^{km}\firsth \right. \\
    & \left. + \frac{a'}{a}\partial_{m}\partial^{k}\firstw D^{m}{_{k}}\firsth  + \frac{a'}{a}\partial^{k}\firstw\partial_{m}D^{m}{_{k}}\firsth - \frac{1}{2}\frac{a'}{a}D^{mk}\firsth D_{km}\dfirsth\right]\delta_{ij} \\
    & -\frac{1}{2}\partial_{i}\partial_{j}\phi^{(2)} + \frac{1}{2}\partial_{i}\partial_{j}\psi^{(2)} - \frac{a'}{a}\partial_{i}\partial_{j}w^{(2)} - \frac{1}{2}\partial_{i}\partial_{j}w^{(2)'} \\
    & - \frac{1}{2}\frac{a'}{a}(\partial_{j}w_{i}^{(2)} + \partial_{i}w_{j}^{(2)}) - \frac{1}{4}(\partial_{j}w_{i}^{(2)'} + \partial_{i}w_{j}^{(2)'}) \\
    & + \frac{1}{2}\left[\frac{a''}{a} +\left(\frac{a'}{a}\right)^{2}\right]\left(D_{ij}h^{(2)}) + \partial_{i}h_{j}^{(2)} + \partial_{j}h^{(2)}{_{i}} + h_{ij}^{(2)}\right) \\
    & + \frac{1}{2}\frac{a'}{a}\left(D_{ij}h^{(2)'}) + \partial_{i}h_{j}^{(2)'} + \partial_{j}h^{(2)'}{_{i}} + h_{ij}^{(2)'}\right)  \\
    & + \frac{1}{4}\left(D_{ij}h^{(2)''}) + \partial_{i}h_{j}^{(2)''} + \partial_{j}h^{(2)''}{_{i}} + h_{ij}^{(2)''}\right) \\
    & + \frac{1}{2}\partial_{k}\partial_{i}D^{k}{_{j}}h^{(2)} - \frac{1}{4}\nabla^{2}h_{ij}^{(2)} - \frac{1}{4}\nabla^{2}D_{ij}h^{(2)} \\
    & + \partial_{i}\firstphi\partial_{j}\firstphi + 2\firstphi\partial_{i}\partial_{j}\firstphi - \partial_{i}\firstphi\partial_{j}\firstpsi  - \partial_{i}\firstpsi\partial_{j}\firstphi + 3\partial_{i}\firstpsi\partial_{j}\firstpsi \\
    & + 2\firstpsi\partial_{i}\partial_{j}\firstpsi  + 4\frac{a'}{a}\firstphi\partial_{i}\partial_{j}\firstw + \dfirstphi\partial_{i}\partial_{j}\firstw + 2\firstphi\partial_{i}\partial_{j}\dfirstw \\
    & - 2\frac{a'}{a}\partial_{i}\firstpsi\partial_{j}\firstw - 2\frac{a'}{a}\partial_{j}\firstpsi\partial_{i}\firstw - \partial_{i}\dfirstpsi\partial_{j}\firstw - \partial_{j}\dfirstpsi\partial_{i}\firstw \\
    & - \partial_{j}\firstpsi\partial_{i}\dfirstw - \partial_{i}\firstpsi\partial_{j}\dfirstw  + \dfirstpsi\partial_{i}\partial_{j}\firstw + \partial_{i}\partial_{j}\firstw\nabla^{2}\firstw \\
    & - \partial_{i}\partial_{k}\firstw\partial_{j}\partial^{k}\firstw  - 2\left[\left(\frac{a'}{a}\right)^{2}+\frac{a''}{a}\right]\firstphi D_{ij}\firsth - 2\frac{a'}{a}\firstphi D_{ij}\dfirsth \\
    & - \frac{a'}{a}\dfirstphi D_{ij}\firsth - \frac{1}{2}\dfirstphi D_{ij}\dfirsth \firstphi D_{ij}\ddfirsth + \frac{1}{2}\partial_{k}\firstphi\partial_{i}D^{k}{_{j}}\firsth \\
    &  + \frac{1}{2}\partial_{k}\firstphi\partial_{j}D^{k}{_{i}}\firsth - \frac{1}{2}\partial_{k}\firstphi\partial^{k}D_{ij}\firsth  + \frac{1}{2}\partial_{k}\firstpsi\partial_{i}D^{k}{_{j}}\firsth + \frac{1}{2}\partial_{k}\firstpsi\partial_{j}D^{k}{_{i}}\firsth \\
    & - \frac{3}{2}\partial_{k}\firstpsi\partial^{k}D_{ij}\firsth + \firstpsi\partial_{k}\partial_{i}D^{k}{_{j}}\firsth + \firstpsi\partial_{k}\partial_{j}D^{k}{_{i}}\firsth  - \firstpsi\nabla^{2}D_{ij}\firsth \\
    & + \partial_{i}\firstpsi\partial_{k}D^{k}{_{j}}\firsth + \partial_{j}\firstpsi\partial_{k}D^{k}{_{i}}\firsth  + \partial_{i}\partial_{k}\firstpsi D^{k}{_{j}}\firsth  + \partial_{j}\partial_{k}\firstpsi D^{k}{_{i}}\firsth \\
    & - 3\frac{a'}{a}\dfirstpsi D_{ij}\firsth + \frac{1}{2}\dfirstpsi D_{ij}\dfirsth + \frac{a'}{a}\partial^{k}\firstw\partial_{i}D_{kj}\firsth  + \frac{a'}{a}\partial^{k}\firstw\partial_{j}D_{ki}\firsth \\
    & - \frac{a'}{a}\partial^{k}\firstw\partial_{k}D_{ij}\firsth - \frac{a'}{a}\nabla^{2}\firstw D_{ij}\firsth + \frac{1}{2}\partial^{k}\dfirstw\partial_{i}D_{kj}\firsth \\
    & + \frac{1}{2}\partial^{k}\dfirstw\partial_{j}D_{ki}\firsth - \frac{1}{2}\partial^{k}\dfirstw\partial_{k}D_{ij}\firsth  + \frac{1}{2}\partial^{k}\firstw\partial_{i}D_{kj}\dfirsth \\
    &  + \frac{1}{2}\partial^{k}\firstw\partial_{j}D_{ki}\dfirsth - \partial^{k}\firstw\partial_{k}D_{ij}\dfirsth  + \frac{1}{2}\partial_{k}\partial_{i}\firstw D^{k}{_{j}}\dfirsth  \\
    &  + \frac{1}{2}\partial_{k}\partial_{j}\firstw D^{k}{_{i}}\dfirsth  - \frac{1}{2}\nabla^{2}\firstw D_{ij}\dfirsth - \frac{1}{2}\partial_{i}D_{mj}\firsth\partial_{k}D^{km}\firsth \\
    & - \frac{1}{2}\partial_{j}D_{mi}\firsth\partial_{k}D^{km}\firsth  + \frac{1}{2}\partial_{m}D_{ij}\firsth\partial_{k}D^{km}\firsth  - \frac{1}{2}\partial_{i}\partial_{k}D_{mj}\firsth D^{km}\firsth \\
    & - \frac{1}{2}\partial_{j}\partial_{k}D_{mi}\firsth D^{km}\firsth  + \frac{1}{2}\partial_{m}\partial_{k}D_{ij}\firsth D^{km}\firsth + \frac{1}{2}\partial_{i}\partial_{j}D_{km}\firsth D^{km}\firsth \\
    & + \frac{1}{4}\partial_{j}D_{km}\firsth\partial_{i}D^{km}\firsth  + \frac{1}{2}\partial_{m}D_{ik}\firsth\partial^{m}D^{k}{_{j}}\firsth \\
    & - \frac{1}{2}\partial_{m}D_{ik}\firsth\partial^{k}D^{m}{_{j}}\firsth - \frac{1}{2}D^{k}{_{i}}\dfirsth D_{kj}\dfirsth. \numberthis
\end{align*}

\newpage

\section{Ricci scalar}\label{ricci_scalar}

The Ricci scalar is a contraction of the Ricci tensor

\begin{equation}
	R = R^{\mu}_{\mu}
\end{equation} 

The components are expanded to the following perturbations, 

\begin{equation}
\label{zeroth_order_R_scalar}
    R^{(0)} = \frac{6}{a^{2}}\frac{a''}{a'}, 
\end{equation}

\begin{align*}
\label{first_order_R_scalar}
    R^{(1)} & = \frac{1}{a^{2}}\left(-2\nabla^{2}\firstphi - 6\ddfirstpsi - 6\frac{a'}{a}\dfirstphi - 18\frac{a'}{a}\dfirstpsi - 12\frac{a''}{a}\firstphi + 4\nabla^{2}\firstpsi \right. \\
    & \left. - 6\frac{a'}{a}\nabla^{2}\firstw - 2\nabla^{2}\dfirstw + \partial^{k}\partial_{m}D^{m}{_{k}}\firsth\right), \numberthis 
\end{align*}

\begin{align*}
\label{second_order_R_scalar}
    R^{(2)} & = \frac{1}{a^{2}}\left(-\nabla^{2}\phi^{(2)} - 3\frac{a'}{a}\phi^{(2)'} - 6\frac{a''}{a}\phi^{(2)} + 2\nabla^{2}\psi^{(2)}-9\frac{a'}{a}\psi^{(2)'} -3\psi^{(2)''}  \right. \\
    & \left. - 3\frac{a'}{a}\nabla^{2}w^{(2)} - \nabla^{2}w^{(2)'} + \frac{1}{2}\partial_{k}\partial_{m}D^{km}h^{(2)} + 24\frac{a''}{a}(\firstphi)^{2} + 2\partial_{k}\firstphi\partial^{k}\firstphi \right. \\
    & \left. + 4\firstphi\nabla^{2}\firstphi  + 24\frac{a'}{a}\firstphi\dfirstphi  + 6\dfirstphi\dfirstpsi + 36\frac{a'}{a}\firstphi\dfirstpsi + 2\partial_{k}\firstpsi\partial^{k}\firstphi \right. \\
    & \left. - 4\firstpsi\nabla^{2}\firstphi + 12\firstphi\ddfirstpsi - 12 \firstpsi\ddfirstpsi - 36\frac{a''}{a}\dfirstpsi\firstpsi + 6\partial^{k}\firstpsi\partial_{k}\firstpsi \right. \\
    & \left.  + 16\firstpsi\nabla^{2}\firstpsi + 6\frac{a'}{a}\partial_{k}\firstphi\partial^{k}\firstw + 12\frac{a'}{a}\firstphi\nabla^{2}\firstw + 4\firstphi\nabla^{2}\dfirstw \right. \\
    & \left. + 2\dfirstphi\nabla^{2}\firstw   + 8\partial^{k}\dfirstpsi\partial_{k}\firstw + 2\partial^{k}\firstpsi\partial_{k}\dfirstw - 4\firstpsi\nabla^{2}\dfirstw \right. \\
    & \left. - 12\frac{a'}{a}\firstpsi\nabla^{2}\firstw  + 4\dfirstpsi\nabla^{2}\firstw  + 6\frac{a'}{a}\partial^{k}\firstpsi\partial_{k}\firstw - 6\frac{a''}{a}\partial_{k}\firstw\partial^{k}\firstw \right. \\
    & \left. - 6\frac{a'}{a}\partial_{k}\firstw\partial^{k}\dfirstw + \nabla^{2}\firstw\nabla^{2}\firstw - \partial_{k}\partial_{m}\firstw\partial^{k}\partial^{m}\firstw  + 2\partial_{k}\firstphi\partial_{m}D^{mk}\firsth \right. \\
    & \left. + 2\partial_{m}\partial_{k}\firstphi D^{km}\firsth  + 4\firstpsi\partial_{k}\partial_{m}D^{mk}\firsth + 2\partial_{m}\partial_{k}\firstpsi D^{km}\firsth  \right. \\
    & \left. + 6\frac{a'}{a}\partial^{k}\firstw\partial_{m}D^{m}{_{k}}\firsth  + 6\frac{a'}{a}\partial_{k}\partial_{m}\firstw D^{km}\firsth + 2\partial_{k}\firstw\partial^{m}D^{k}{_{m}}\dfirsth \right. \\
    & \left.  + 2\partial_{m}\dfirstw\partial_{k}D^{km}\firsth  + 2\partial_{k}\partial_{m}\dfirstw D^{km}\firsth  + \partial_{k}\partial_{m}\firstw D^{km}\dfirsth \right. \\
    & \left. - 2\partial_{k}\partial^{m}D_{nm}\firsth D^{kn}\firsth - \partial_{k}D^{km}\firsth\partial^{n}D_{mn}\firsth + \nabla^{2}D_{km}\firsth D^{mn}\firsth \right. \\
    & \left. + \frac{3}{4}\partial^{n}D^{km}\firsth\partial_{n}D_{km}\firsth  - \frac{1}{2}\partial_{k}D_{mn}\firsth\partial^{n}D^{mk}\firsth - 3\frac{a'}{a}D^{km}\firsth D_{km}\dfirsth \right. \\
    & \left. - \frac{3}{4}D^{km}\dfirsth D_{km}\dfirsth - D^{km}\firsth D_{km}\ddfirsth\right). \numberthis
\end{align*}

\section{Einstein tensor}\label{einstein_tensor}

\begin{equation}
	G^{\lambda}_{\mu} = g^{\nu\lambda} \left(R_{\mu\nu} - \frac{1}{2} g_{\mu\nu}R\right)
\end{equation}

\subsection{Background components}

\begin{align*}
    G^{(0)0}{_{0}} &= -\frac{3}{a^{2}}\left(\frac{a'}{a}\right)^{2} , \numberthis \\
    G^{(0)0}{_{i}} &= G^{(0)i}{_{0}} = 0, \numberthis\\
    G^{(0)i}{_{j}} &= -\frac{1}{a^{2}}\left[2\frac{a''}{a}-\left(\frac{a'}{a}\right)^{2}\right]\delta^{i}{_{j}}. \numberthis
\end{align*}

\subsection{First-order components}

\begin{align*}
    G^{(1)0}{_{0}} &= \frac{1}{a^{2}}\left[6\left(\frac{a'}{a}\right)^{2}\firstphi + 6\frac{a'}{a}\dfirstpsi - 2\nabla^{2}\firstpsi + 2\frac{a'}{a}\nabla^{2}\firstw  - \frac{1}{2}\partial^{k}\partial_{m}D^{m}{_{k}}\firsth\right] , \numberthis \\
    G^{(1)0}{_{i}} & = \frac{1}{a^{2}}\left[-2\frac{a'}{a}\partial_{i}\firstphi - 2\partial_{i}\dfirstpsi - \frac{1}{2}\partial_{k}D^{k}{_{i}}\dfirsth\right], \numberthis\\
    G^{(1)i}{_{0}} &= \frac{1}{a^{2}}\left[4\left(\frac{a'}{a}\right)^{2}\partial^{i}\firstw - 2\frac{a''}{a}\partial^{i}\firstw + 2\partial^{i}\dfirstpsi + 2\frac{a'}{a}\partial^{i}\firstphi + \frac{1}{2}\partial_{k}D^{ki}\dfirsth\right], \numberthis\\
    G^{(1)i}{_{j}} &= \frac{1}{a^{2}}\left[\left(\left[4\frac{a''}{a}-2\left(\frac{a'}{a}\right)^{2}\right]\firstphi+2\frac{a'}{a}\dfirstphi + \nabla^{2}\firstphi  + 4\frac{a'}{a}\dfirstpsi \right. \right. \\
    & \left. \left. + 2\ddfirstpsi - \nabla^{2}\firstpsi  + 2\frac{a'}{a}\nabla^{2}\firstw + \nabla^{2}\dfirstw - \frac{1}{2}\partial_{k}\partial^{m}D^{k}{_{m}}\firsth\right)\delta^{i}{_{j}} \right. \\
    & \left. - \partial^{i}\partial_{j}\firstphi + \partial^{i}\partial_{j}\firstpsi - 2\frac{a'}{a}\partial^{i}\partial_{j}\firstw - \partial^{i}\partial_{j}\dfirstw  + \frac{a'}{a}D^{i}{_{j}}\dfirsth \right. \\
    & \left. + \frac{1}{2}D^{i}{_{j}}\ddfirsth + \frac{1}{2}\partial^{i}\partial_{k}D^{k}{_{j}}\firsth + \frac{1}{2}\partial_{k}\partial_{j}D^{ik}\firsth - \frac{1}{2}\partial^{k}\partial_{k}D^{i}{_{j}}\firsth\right]. \numberthis
\end{align*}

\subsection{Second-order components}

\begin{align*}
    G^{(2)0}{_{0}} &= \frac{1}{a^{2}}\left[3\left(\frac{a'}{a}\right)^{2}\phi^{(2)} - \nabla^{2}\psi^{(2)} + 3\frac{a'}{a}\psi^{(2)'} + \frac{a'}{a}\nabla^{2}w^{(2)} - \frac{1}{4}\partial_{k}\partial_{m}D^{km}h^{(2)} \right. \\
    & \left. - 12\left(\frac{a'}{a}\right)^{2}(\firstphi)^{2} +3\left(\frac{a'}{a}\right)^{2}\partial_{k}\firstw\partial^{k}\firstw - 12\frac{a'}{a}\firstphi\dfirstpsi - 3\partial^{k}\firstpsi\partial_{k}\firstpsi \right. \\
    & \left. - 8\firstpsi\nabla^{2}\firstpsi + 12\frac{a'}{a}\firstpsi\dfirstpsi - 3(\firstpsi)^{2} - 4\frac{a'}{a}\firstphi\nabla^{2}\firstw \right. \\
    & \left. - 2\frac{a'}{a}\partial^{k}\firstphi\partial_{k}\firstw - 2\frac{a'}{a}\partial_{k}\firstpsi\partial^{k}\firstw + 4\frac{a'}{a}\firstpsi\nabla^{2}\firstw - 2\partial^{k}\dfirstpsi\partial_{k}\firstw \right. \\
    & \left. - 2\dfirstpsi\nabla^{2}\firstw + \frac{1}{2}\partial_{k}\partial_{m}\firstw\partial^{k}\partial^{m}\firstw - \frac{1}{2}\nabla^{2}\firstw\nabla^{2}\firstw \right. \\
    & \left. - 2\firstpsi\partial_{k}\partial^{m}D^{k}{_{m}}\firsth + \partial_{k}\partial_{m}\firstpsi D^{km}\firsth - 2\frac{a'}{a}\partial_{k}\partial_{m}\firstw D^{km}\firsth \right. \\
    & \left. - \frac{1}{2}\partial_{k}\partial_{m}\firstw D^{km}\dfirsth - 2\frac{a'}{a}\partial_{k}\firstw\partial_{m}D^{km}\firsth - \frac{1}{2}\partial_{k}\firstw\partial_{m}D^{km}\dfirsth \right. \\
    & \left.  - \frac{1}{2}\nabla^{2}D_{mk}\firsth D^{mk}\firsth  + \partial_{n}\partial^{k}D_{mk}\firsth D^{mn}\firsth  + \frac{1}{2}\partial_{k}D^{km}\firsth\partial^{n}D_{mn}\firsth \right. \\
    & \left.  - \frac{3}{8}\partial^{n}D^{km}\firsth\partial_{n}D_{km}\firsth + \frac{1}{4}\partial_{k}D_{mn}\firsth\partial^{m}D^{kn}\firsth \right. \\
    & \left. + \frac{a'}{a}D^{kn}\firsth D_{kn}\dfirsth + \frac{1}{8}D^{kn}\dfirsth D_{kn}\dfirsth \right], \numberthis
\end{align*}

\begin{align*}
    G^{(2)0}{_{i}} & = \frac{1}{a^{2}}\left[-\frac{a'}{a}\partial_{i}\phi^{(2)} - \partial_{i}\psi^{(2)'} + \frac{1}{4}\nabla^{2}w_{i}^{(2)} - \frac{1}{4}\partial_{k}D^{k}{_{i}}h^{(2)'}-\frac{1}{4}\nabla^{2}h_{i}^{(2)'}\right. \\
    & \left. 8\frac{a'}{a}\firstphi\partial_{i}\firstphi + 4\firstphi\partial_{i}\dfirstpsi + 2\partial_{i}\firstphi\dfirstpsi - 4\dfirstpsi\partial_{i}\firstpsi - 4\dfirstpsi\partial_{i}\dfirstpsi \right. \\
    & \left. + \partial_{i}\firstphi\nabla^{2}\firstw - \partial^{k}\firstphi\partial_{i}\partial_{k}\firstw  + \nabla^{2}\firstpsi\partial_{i}\firstw \right. \\
    & \left. + \partial_{i}\partial_{k}\firstpsi\partial^{k}\firstw - 2\frac{a'}{a}\partial_{k}\partial_{i}\firstw\partial^{k}\firstw + \firstphi\partial^{k}D_{ki}\dfirsth \right. \\
    & \left. + \frac{1}{2}\partial^{k}\firstphi D_{ik}\dfirsth -\firstpsi\partial_{k}D^{k}{_{i}}\dfirsth  + \frac{1}{2}\partial_{k}\firstpsi D^{k}{_{i}}\dfirsth  \right. \\
    & \left. - \dfirstpsi\partial_{k}D^{k}{_{i}}\firsth - \partial_{k}\dfirstpsi D^{k}{_{i}}\firsth + \frac{1}{2}\partial_{k}\firstw\partial^{k}\partial^{m}D_{im}\firsth \right. \\
    & \left. - \frac{1}{2}\partial^{m}\firstw\nabla^{2}D_{im}\firsth  + \frac{1}{2}\partial_{k}D^{km}\firsth D_{im}\dfirsth + \frac{1}{2}D^{km}\firsth\partial_{k}D_{im}\dfirsth \right. \\
    & \left. - \frac{1}{4}\partial_{i}D_{mk}\firsth D^{km}\dfirsth - \frac{1}{2}D^{km}\firsth\partial_{i}D_{mk}\dfirsth\right], \numberthis
\end{align*}

\begin{align*}
    G^{(2)i}{_{0}} &= \frac{1}{a^{2}}\left[\frac{a'}{a}\partial^{i}\phi^{(2)} + \partial^{i}\psi^{(2)'} - \frac{1}{4}\nabla^{2}w^{i(2)} + \left(2\left(\frac{a'}{a}\right)^{2}- \frac{a''}{a}\right)\partial^{i}w^{(2)} \right. \\
    & \left. + \left(2\left(\frac{a'}{a}\right)^{2} - \frac{a''}{a}\right)w^{i(2)}+ \frac{1}{4}\partial_{k}D^{ki}h^{(2)'} + \frac{1}{4}\nabla^{2}h^{i(2)'}  \right. \\
    & \left. - 4\frac{a'}{a}\firstphi\partial^{i}\firstphi + 4\frac{a'}{a}\partial^{i}\firstphi\firstpsi - 2\partial^{i}\firstphi\dfirstpsi + 4\dfirstpsi\partial^{i}\firstpsi + 8\partial^{i}\dfirstpsi\firstpsi \right. \\
    & \left. - \partial^{i}\firstphi\nabla^{2}\firstw - \partial^{i}\partial_{k}\firstphi\partial^{k}\firstw  + \nabla^{2}\firstphi\partial^{i}\firstw  + \partial^{k}\firstphi\partial^{i}\partial_{k}\firstw \right. \\
    & \left. + \left(4\frac{a''}{a}-8\left(\frac{a'}{a}\right)^{2}\right)\firstphi\partial^{i}\firstw + 2\frac{a'}{a}\dfirstphi\partial^{i}\firstw + 2\ddfirstphi\partial^{i}\firstw \right. \\
    & \left. + \left(8\left(\frac{a'}{a}\right)^{2}-4\frac{a''}{a}\right)\firstpsi\partial^{i}\firstw  - 2\frac{a'}{a}\dfirstpsi\partial^{i}\firstw + \nabla^{2}\dfirstw\partial^{i}\firstw \right. \\
    & \left. - \partial^{i}\partial_{k}\dfirstw\partial^{k}\firstw  - \frac{1}{2}\partial^{k}\firstphi D^{i}{_{k}}\dfirsth  - 2\frac{a'}{a}\partial_{k}\firstphi D^{ki}\firsth - \frac{1}{2}\partial_{k}\firstpsi D^{ki}\dfirsth \right. \\
    & \left. + 2\firstpsi\partial_{k}D^{ki}\dfirsth + \dfirstpsi\partial_{k}D^{ki}\firsth - \partial_{k}\dfirstpsi D^{ki}\firsth \right. \\
    & \left. + \left(2\frac{a''}{a}-4\left(\frac{a'}{a}\right)^{2}\right)\partial_{k}\firstw D^{ik}\firsth + \frac{a'}{a}\partial^{k}\firstw D^{i}{_{k}}\dfirsth+ \frac{1}{2}\partial^{k}\firstw D^{i}{_{k}}\ddfirsth  \right. \\
    & \left. - \frac{1}{2}\partial_{k}D^{km}\firsth D^{i}{_{m}}\dfirsth - \frac{1}{2}D^{km}\firsth \partial_{k}D^{i}{_{m}}\dfirsth + \frac{1}{4}\partial^{i}D_{mk}\firsth D^{km}\dfirsth \right. \\
    & \left. + \frac{1}{2}D^{km}\firsth\partial^{i}D_{mk}\dfirsth - \frac{1}{2}D^{ik}\firsth\partial_{m}D^{m}{_{k}}\dfirsth\right], \numberthis
\end{align*}

\begin{align*}
    G^{d,i(2)}{_{j}} & = a^{-2}\left[\frac{1}{2}\nabla^{2}\phi^{(2)}+\left[2\frac{a''}{a}-\left(\frac{a'}{a}\right)^{2}\right]\phi^{(2)} + \frac{a'}{a}\phi^{(2)'}-\frac{1}{2}\nabla^{2}\psi^{(2)}+\psi^{(2)''} \right. \\
    & \left. + 2\frac{a'}{a}\psi^{(2)} + \frac{a'}{a}\nabla^{2}w^{(2)} + \frac{1}{2}\nabla^{2}w^{(2)'}  - \frac{1}{4}\partial_{k}\partial_{m}D^{km}h^{(2)} \right. \\
    & \left. + \left[4\left(\frac{a'}{a}\right)^{2}-8\frac{a''}{a}\right](\firstphi)^{2}-8\frac{a'}{a}\firstphi\dfirstphi - \partial_{k}\firstphi\partial^{k}\firstphi - 2\firstphi\nabla^{2}\firstphi \right. \\
    & \left. - 4\firstphi\ddfirstpsi - 2\dfirstphi\dfirstpsi  - 8\frac{a'}{a}\firstphi\dfirstpsi  -2\partial_{k}\firstpsi\partial^{k}\firstpsi \right. \\
    & \left. - 4\firstpsi\nabla^{2}\firstpsi + (\dfirstpsi)^{2} + 8\frac{a'}{a}\firstpsi\dfirstpsi + 4\firstpsi\ddfirstpsi \right. \\
    & \left. + 2\firstpsi\nabla^{2}\firstphi - \dfirstphi\nabla^{2}\firstw - 2\firstphi\nabla^{2}\dfirstw - 2\frac{a'}{a}\partial_{k}\firstw\partial^{k}\dfirstw \right. \\
    & \left. - \frac{1}{2}\nabla^{2}\firstw\nabla^{2}\firstw + \frac{1}{2}\partial^{m}\partial^{k}\firstw\partial_{m}\partial_{k}\firstw + 4\frac{a'}{a}\firstpsi\nabla^{2}\firstw + 2\firstpsi\nabla^{2}\dfirstw \right. \\
    & \left. - 2\partial^{k}\dfirstpsi\partial_{k}\firstw - \dfirstpsi\nabla^{2}\firstw - \partial_{k}\partial_{m}\firstphi D^{km}\firsth - \partial_{k}\firstphi\partial_{m}D^{km}\firsth \right. \\
    & \left. - \partial_{k}\firstpsi\partial_{m}D^{km}\firsth - \partial_{k}\firstw\partial^{m}D^{k}{_{m}}\dfirsth  - \partial_{k}\dfirstw\partial_{m}D^{km}\firsth \right. \\
    & \left. + \partial_{k}\partial_{m}\dfirstw D^{km}\firsth - 2\frac{a'}{a}\partial^{k}\firstw\partial_{m}D^{m}{_{k}}\dfirsth - 2\frac{a'}{a}\partial^{k}\partial_{m}\firstw D^{m}{_{k}}\firsth \right. \\
    & \left. -\frac{1}{2}\partial_{k}\partial_{m}\firstw D^{km}\dfirsth - 2\firstpsi\partial_{k}\partial_{m}D^{km}\firsth + \partial_{k}\partial^{n}D_{mn}\firsth D^{km}\firsth \right. \\
    & \left. - \frac{1}{2}\nabla^{2}D_{mn}\firsth D^{mn}\firsth + \frac{1}{2}\partial_{k}D^{k}{_{m}}\firsth\partial^{n}D_{n}^{m}\firsth + \frac{1}{2}D^{mk}\firsth D_{mk}\ddfirsth  \right. \\
    & \left. - \frac{3}{8}\partial^{n}D_{km}\firsth\partial^{n}D^{km}\firsth + \frac{3}{8}D^{mk}\dfirsth D_{mk}\dfirsth \right. \\
    & \left.  + \frac{a'}{a}D^{mk}\firsth D_{mk}\dfirsth + \frac{1}{4}\partial^{n}D^{km}\firsth\partial_{m}D_{kn}\firsth\right]\delta^{i}{_{j}}, 
\end{align*}

\begin{align*}
\label{non_diagonal_second_order_G}
    G^{nd,i(2)}{_{j}} &= a^{-2}\left[-\frac{1}{2}\partial^{i}\partial_{j}\phi^{(2)} + \frac{1}{2}\partial^{i}\partial_{j}\psi^{(2)} - \frac{a'}{a}\partial^{i}\partial_{j}w^{(2)} - \frac{1}{2}\partial^{i}\partial_{j}w^{(2)'} \right. \\
    & \left. -\frac{1}{2}\frac{a'}{a}\left(\partial^{i}w_{j}^{(2)} + \partial_{j}w^{i(2)}\right) - \frac{1}{4}\left(\partial^{i}w_{j}^{(2)'} + \partial_{j}w^{i(2)'}\right) \right. \\
    & \left. + \frac{1}{2}\frac{a'}{a}\left(D^{i}{_{j}}h^{(2)'}+\partial^{i}h_{j}^{(2)'} + \partial_{j}h^{i(2)'} + \dtensorh\right) + \frac{1}{2}\partial_{k}\partial^{i}D^{k}{_{j}}h^{(2)} \right. \\
    & \left. - \frac{1}{4}\nabla^{2}D^{i}{_{j}}h^{(2)} - \frac{1}{4}\nabla^{2}\tensorh \right. \\
    & \left. + \frac{1}{4}\left(D^{i}{_{j}}h^{(2)''} + +\partial^{i}h_{j}^{(2)''} + \partial_{j}h^{i(2)''} + \ddtensorh\right)\right.\\
    & \left. + \partial^{i}\firstphi\partial_{j}\firstphi + 2\firstphi\partial^{i}\partial_{j}\firstphi - 2\firstpsi\partial^{i}\partial_{j}\firstphi - \partial_{j}\firstphi\partial^{i}\firstpsi \right. \\
    & \left. - \partial^{i}\firstphi\partial_{j}\firstpsi  + 3\partial^{i}\firstpsi\partial_{j}\firstpsi + 4\firstpsi\partial^{i}\partial_{j}\firstpsi + 2\frac{a'}{a}\partial^{i}\firstw\partial_{j}{(\firstphi} \right. \\
    & \left.  + 4\frac{a'}{a}\firstphi\partial^{j}\partial_{i}\firstw  + \dfirstphi\partial^{i}\partial_{j}\firstw + 2\firstphi\partial^{i}\partial_{j}\dfirstw + \nabla^{2}\firstw\partial^{i}\partial_{j}\firstw \right. \\
    & \left. - \partial_{j}\partial^{k}\firstw\partial^{i}\partial_{k}\firstw  - 2\frac{a'}{a}\partial^{i}\firstpsi\partial_{j}\firstw - 2\frac{a'}{a}\partial^{i}\firstw\partial_{j}\firstpsi - \partial^{i}\dfirstpsi\partial_{j}\firstw \right. \\
    & \left. + \frac{a'}{a}\partial^{i}\firstw\partial_{j}\dfirstpsi  - \partial^{i}\firstpsi\partial_{j}\dfirstw - \frac{a'}{a}\partial^{i}\dfirstw\partial_{j}\firstpsi - 2\firstpsi\partial^{i}\partial_{j}\dfirstw \right. \\
    & \left. + \dfirstpsi\partial^{i}\partial_{j}\firstw  - 2\frac{a'}{a}\firstpsi\partial^{i}\partial_{j}\firstw - 2\frac{a'}{a}\firstphi D^{i}{_{j}}\dfirsth - \frac{1}{2}\dfirstphi D^{i}{_{j}}\dfirsth \right. \\
    & \left. - \firstphi D^{i}{_{j}}\ddfirsth  + \frac{1}{2}\partial_{k}\firstphi\partial^{i}D^{k}{_{j}}\firsth + \frac{1}{2}\partial_{k}\firstphi\partial_{j}D^{ki}\firsth - \frac{1}{2}\partial_{k}\firstphi\partial^{k}D^{i}{_{j}}\firsth \right. \\
    & \left. + \partial_{j}\partial_{k}\firstphi D^{ki}\firsth  + \frac{1}{2}\dfirstpsi D^{i}{_{j}}\dfirsth + \ddfirstpsi D^{i}{_{j}}\firsth + 2\frac{a'}{a}\dfirstpsi D^{i}{_{j}}\firsth \right. \\
    & \left. + \frac{1}{2}\partial_{k}\firstpsi\partial^{i}D^{k}{_{j}}\firsth + 2\frac{a'}{a}\firstpsi D^{i}{_{j}}\dfirsth + \firstpsi D^{i}{_{j}}\ddfirsth + \frac{1}{2}\partial_{k}\firstpsi\partial_{j}D^{ki}\firsth \right. \\
    & \left.  - \frac{3}{2}\partial_{k}\firstpsi\partial^{k}D^{i}{_{j}}\firsth  + 2\firstpsi\partial_{k}\partial^{i}D^{k}{_{j}}\firsth + 2\firstpsi\partial_{k}\partial_{j}D^{ki}\firsth \right. \\
    & \left. - 2\firstpsi\nabla^{2}D^{i}{_{j}}\firsth - \nabla^{2}\firstpsi D^{i}{_{j}}\firsth + \partial^{i}\firstpsi\partial_{k}D^{k}{_{j}}\firsth \right. \\
    & \left. + \partial_{j}\firstpsi\partial_{k}D^{ki}\firsth + \partial_{k}\partial^{i}\firstpsi D^{k}{_{j}}\firsth + \frac{1}{2}\partial^{i}\firstw\partial_{k}D^{k}{_{j}}\dfirsth \right. \\
    & \left. + \frac{1}{2}\partial_{k}\partial^{i}\firstw D^{k}{_{j}}\dfirsth + \frac{1}{2}\partial_{k}\partial_{j}\firstw D^{ki}\dfirsth - \frac{1}{2}\nabla^{2}\firstw D^{i}{_{j}}\dfirsth \right. \\
    & \left. + \frac{1}{2}\partial^{k}\firstw\partial^{i}D_{kj}\dfirsth + \frac{1}{2}\partial^{k}\firstw\partial_{j}D^{i}{_{k}}\dfirsth \right. \\
    & \left. - \partial^{k}\firstw\partial_{k}D^{i}{_{j}}\dfirsth + \frac{1}{2}\partial^{k}\dfirstw\partial^{i}D_{kj}\firsth + \frac{1}{2}\partial^{k}\dfirstw\partial_{j}D^{i}{_{k}}\firsth \right. \\
    & \left. - \frac{1}{2}\partial^{k}\dfirstw\partial_{k}D^{i}{_{j}}\firsth + \partial_{k}\partial_{j}\dfirstw D^{ik}\firsth + \frac{a'}{a}\partial^{k}\firstw\partial^{i}D_{kj}\firsth \right. \\
    & \left. + \frac{a'}{a}\partial^{k}\firstw\partial_{j}D^{i}{_{k}}\firsth - \frac{a'}{a}\partial^{k}\firstw\partial_{k}D^{i}{_{j}}\firsth + 2\frac{a'}{a}\partial_{k}\partial_{j}\firstw D^{ik}\firsth \right. \\
    & \left. - \frac{1}{2}D^{ki}\dfirsth D_{kj}\dfirsth  - \frac{1}{2}\partial^{i}D_{mj}\firsth\partial_{k}D^{km}\firsth - \frac{1}{2}\partial_{j}D^{i}{_{m}}\firsth\partial_{k}D^{km}\firsth \right. \\
    & \left.+ \frac{1}{2}\partial_{m}D^{i}{_{j}}\firsth\partial_{k}D^{km}\firsth  - \frac{1}{2}\partial_{k}\partial^{i}D_{mj}\firsth D^{km}\firsth \right. \\
    & \left.- \frac{1}{2}\partial_{k}\partial_{j}D^{i}{_{m}}\firsth D^{km}\firsth + \frac{1}{2}\partial_{k}\partial_{m}D^{i}{_{j}}\firsth D^{km}\firsth \right. \\
    & \left. + \frac{1}{2}D^{km}\firsth\partial^{i}\partial_{j}D_{km}\firsth + \frac{1}{4}\partial^{i}D^{mk}\firsth\partial_{j}D_{mk}\firsth \right. \\
    & \left. - \frac{a'}{a}D_{kj}\dfirsth D^{ik}\firsth - \frac{1}{2}D_{kj}\ddfirsth D^{ki}\firsth \right. \\
    & \left. - \partial_{m}\partial_{k}D^{m}{_{j}}\firsth D^{ki}\firsth + \frac{1}{2}\partial_{m}\partial^{m}D_{kj}\firsth D^{ki}\firsth \right. \\
    & \left. + \frac{1}{2}\partial_{m}D^{ik}\firsth\partial^{m}D_{kj}\firsth - \frac{1}{2}\partial_{m}D^{ik}\firsth\partial_{k}D^{m}{_{j}}\firsth \right]. \numberthis   
\end{align*}

\newpage

\section{Energy-momentum tensor}\label{energy_momentum_tensor}

The energy-momentum tensor describes the perturbed matter quantities. The following is the expression for a fluid, 

\begin{equation}
	T^{\mu}_{\nu} = (\rho + p)u^{\mu}u_{\nu} + p\delta^{\mu}_{\nu} + \pi^{\mu}_{\nu}
\end{equation}

where the perturbed energy density, $\rho$, and pressure, $p$. 

\begin{equation}
    \rho = \bar{\rho} + \sum_{r=1}^{\infty}\frac{1}{r!} \delta^{(r)}\rho, 
\end{equation}
\begin{equation}
    P = \bar{P} + \sum_{r=1}^{\infty}\frac{1}{r!} \delta^{(r)}P.
\end{equation}

The 4 - velocity, $u^{\mu}$, can be defined in the following way, 

\begin{equation}
    u^{\mu} = \frac{1}{a}\left(\delta^{\mu}_0 + \sum_{r=1}^{\infty}\frac{1}{r!}v^{\mu}_{(r)}\right), 
\end{equation}

\begin{equation}
    u^{0} = \frac{1}{a}\left(1 + v^{(1)0} + \frac{1}{2}v^{(2)0}\right), 
\end{equation}
\begin{equation}
    u^{i} = \frac{1}{a}\left(v^{(1)i} + \frac{1}{2}v^{(2)i}\right), 
\end{equation}

where $u^{\mu}$ is subject to normalization condition $u^{\mu}u^{\nu}g_{\mu\nu} = -1$. Using this condition, the components of 4-velocity, $u^{0}$, $u^{i}$, $u_{0}$ and $u_{i}$, up to second order can be found. First by finding, $v^{(1)0}$ and $v^{(2)0}$. For first and second order perturbation, 

\begin{align*}
    u^{0}u^{0}g_{00} & = -1 \\
    \left(1+v^{(1)0}\right)^{2}\left(-1-2\phi^{(1)}\right) & = -1 \\
    v^{(1)0} & = -\phi^{(1)} \numberthis
\end{align*}

\begin{equation*}
    u^{0}u^{0}g_{00} + 2u^{0}u^{i}g_{0i} + u^{i}u^{j}g_{ij}= -1
\end{equation*}
\begin{align*}
    &\left(1+v^{(1)0} + \frac{1}{2}v^{(2)0}\right)^{2}\left(-1-2\phi^{(1)} - \phi^{(2)}\right)  \\ 
    & + 2\left(1+v^{(1)0} + \frac{1}{2}v^{(2)0}\right)\left(v^{(1)i} + \frac{1}{2}v^{(2)i}\right)\left(\partial_{i}\firstw + \frac{1}{2}\partial_{i}w^{(2)} + \frac{1}{2}w_{i}^{(2)}\right) \\
    & + \left(v^{(1)i} + \frac{1}{2}v^{(2)i}\right)\left(v^{(1)j} + \frac{1}{2}v^{(2)j}\right)\left[\left(1-2\psi^{(1)} - \psi^{(2)}\right)\delta_{ij} \right. \\
    & \left. + D_{ij}\left(h^{(1)} + \frac{1}{2}h^{(2)}\right) + \frac{1}{2}\left(\partial_{i}h_{(2)j} + \partial_{j}h_{(2)i} + h^{(2)}_{ij} \right)\right] = -1,    
\end{align*}

substituting in the expression for $v^{(1)0}$ we get, 

\begin{equation}
    v^{0(2)} = -\phi^{(2)} + 3\left(\phi^{(1)}\right)^{2} + v^{(1)i}v_{i(1)} + 2v^{(1)i}\partial_{i}\firstw. 
\end{equation}

By using the condition $u^{\mu}u_{\mu} = -1$, all components of the 4-velocity is found to be the following, 

\begin{align*}
    u^{0} & = a^{-1}\left[1 - \firstphi - \frac{1}{2}\phi^{(2)} + \frac{3}{2}\left(\firstphi\right)^{2} + \frac{1}{2}v^{(1)k}v_{(1)k} + v_{k}^{(1)}\partial^{k}\firstw\right], \\
    u^{i} & = a^{-1}\left[v^{i(1)} + \frac{1}{2}v^{i(2)}\right], \numberthis \\
    u_{0} & = a\left[-1 - \firstphi - \frac{1}{2}\phi^{(2)} + \frac{1}{2}\left(\firstphi\right)^{2} - \frac{1}{2}v^{i(1)}v_{i(1)}\right], \\
    u_{i} & = a\left[\partial_{i}\firstw - v_{i}^{(1)} + \frac{1}{2}\left(\partial_{i}w^{(2)} + w_{i}^{(2)} + v_{i}^{(2)}\right) \right. \\
    & \left. - \partial_{i}\firstw\firstphi - 2v^{k(1)}\firstpsi + v^{k(1)}D_{ik}\firsth \right], \numberthis
\end{align*}

The anisotropic stress tensor, $\pi_{\mu\nu}$, can also be split into first and second order parts, 

\begin{equation}
    \pi_{\mu\nu} = \pi^{(1)}_{\mu\nu} + \frac{1}{2}\pi^{(2)}_{\mu\nu}, 
\end{equation}

and it is subject to the constraint $\pi^{\mu}_{\mu}=0$ and $\pi_{\mu\nu}u^{\mu}=0$. The anisotropic stress tensor decomposes into trace-free scalar part,$\Pi$, vector part $\Pi_{i}$, and tensor, $\Pi_{ij}$ part \cite{Malik:2008im}. 

\begin{equation}
    \pi_{ij} = a^{2} \left[\Pi_{,ij} - \frac{1}{3}\delta_{ij}\nabla^{2}\Pi + \frac{1}{2}\left(\Pi_{i,j} + \Pi_{j,i}\right) + \Pi_{ij}\right]
\end{equation}

From the definitions above, the components of the stress energy tensor can be found. 

\subsection{Time - time component of energy-momentum tensor}

\begin{align*}
    T^{0}{_{0}} & = T^{(0)0}{_{0}} + T^{(1)0}{_{0}} + T^{(2)0}{_{0}}, \\
     &= (\rho + p)u^{0}u_{0} + p\delta^{0}{_{0}} + \pi^{0}{_{0}}, \\
    & = \left(\bar{\rho} + \rho^{(1)} + \frac{1}{2}\rho^{(2)} + \bar{P} + P^{(1)} + \frac{1}{2}P^{(2)}\right)\left(-v^{k(1)}v_{k(1)} - \partial^{k}\firstw v_{k}^{(1)}\right) \\
    & - \left(\bar{\rho} + \rho^{(1)} + \frac{1}{2}\rho^{(2)}\right).\numberthis
\end{align*}

\begin{equation}
    T^{(0)0}{_{0}} = -\bar{\rho},
\end{equation}
\begin{equation}
    T^{(1)0}{_{0}} = -\rho^{(1)},
\end{equation}
\begin{equation}
    T^{0(2)}{_{0}} = -\rho^{(2)} - 2\left(\bar{\rho} + \bar{P}\right)v_{k}^{(1)}\left(v^{k(1)} + \partial^{k}\firstw\right).
\end{equation}

\subsection{Time - space component of energy-momentum tensor}

\begin{align*}
    T^{0}{_{i}} & = T^{(0)0}{_{i}} + T^{(1)0}{_{i}} + T^{(2)0}{_{i}}, \\
    & = (\rho + p)u^{0}u_{i} + p\delta^{0}{_{i}} + \pi^{0}{_{i}}, \\
    & = \left(\bar{\rho} + \rho^{(1)} + \frac{1}{2}\rho^{(2)} + \bar{P} + P^{(1)} + \frac{1}{2}P^{(2)}\right) \left(\partial_{i}\firstw + v_{i}^{(1)} \right. \\
    & \left. + \frac{1}{2}\left(\partial_{i}w^{(2)} + w_{i}^{(2)} + v_{i}^{(2)}\right)  - v_{i}^{(1)}\firstphi -2\partial_{i}\firstw\firstphi - 2v^{k(1)}\firstpsi + v^{k(1)}D_{ik}\firsth \right) \\
    & + 2a^{-2}\pi^{(1)}_{ik}\left(\partial^{k}\firstw + v^{k(1)}\right). \numberthis
\end{align*}

\begin{equation}
    T^{(0)0}{_{i}} = 0,
\end{equation}
\begin{equation}
    T^{(1)0}{_{i}} = \left(\bar{\rho} + \bar{P}\right)\left(\partial_{i}\firstw + v_{i}^{(1)}\right),
\end{equation}
\begin{align*}
    T^{(2)0}{_{i}} &= \left(\bar{\rho} + \bar{P}\right)\left[\partial_{i}w^{(2)} + w_{i}^{(2)} + v_{i}^{(2)} - 2\firstphi\left(v_{i}^{(1)} + 2\partial_{i}\firstw\right) \right. \\
    & \left. - 4v^{k(1)}\firstpsi\delta_{ik} + 2v^{k(1)}D_{ik}\firsth\right] \\
    & + 2\left(\rho^{(1)} + P^{(1)}\right)\left(\partial_{i}\firstw + v_{i}^{(1)}\right) + a^{-2}\pi^{(1)}_{ik}\left(\partial^{k}\firstw + v^{k(1)}\right). \numberthis
\end{align*}

\subsection{Space - space component of energy-momentum tensor}

\begin{align*}
    T^{i}{_{j}} & = T^{(0)i}{_{j}} + T^{(1)i}{_{j}} + T^{(2)i}{_{j}}, \\
    & = (\rho + p)u^{i}u_{j} + p\delta^{i}{_{j}} + \pi^{i}{_{j}}, \\
    & = \left(\bar{\rho} + \rho^{(1)} + \frac{1}{2}\rho^{(2)} + \bar{P} + P^{(1)} + \frac{1}{2}P^{(2)}\right) \left(v^{i(1)}\partial_{j}\firstw + v^{i(1)}v_{j(1)}\right) \\
    & + \left(\bar{P} + P^{(1)} + \frac{1}{2}P^{(2)}\right)\delta^{i}{_{j}} + a^{-2}\pi^{(1)i}{_{j}} \\
    & + \frac{1}{2}a^{-2}\left[\pi^{(2)i}{_{j}} + \left(4\firstpsi\delta^{ik} - 2D^{ik}\firsth\right)\pi^{(1)}_{jk}\right]. \numberthis
\end{align*}

\begin{equation}
    T^{(0)i}{_{j}} = \bar{P}\delta^{i}{_{j}},
\end{equation}
\begin{equation}
    T^{(1)i}{_{j}} = P^{(1)}\delta^{i}{_{j}} + a^{-2}\pi^{(1)i}{_{j}},
\end{equation}
\begin{align*}
\label{energy-momentum_second}
    T^{(2)i}{_{j}} &= P^{(2)}\delta^{i}{_{j}} + 2\left(\bar{\rho} + \bar{P}\right)v^{i(1)}\left(\partial_{j}\firstw + v_{j(1)}\right) \\
    & + a^{-2}\left[\pi^{(2)i}{_{j}} + \left(4\firstpsi\delta^{ik} - 2D^{ik}\firsth\right)\pi^{(1)}_{jk}\right]. \numberthis
\end{align*}

\section{Projected spatial Einstein- and energy momentum tensor}\label{projected}

Here we present the spatial Einstein tensor \cref{non_diagonal_second_order_G}, and energy-momentum tensor, \cref{energy-momentum_second}, with the projection tensor, $P^{li}_{jm}$ \cite{Carbone:2004iv}, applied,
\begin{align*}
\label{second_order_G_simplified}
    P^{li}_{jm}G^{(2)m}{_{l}} &= a^{-2}\left[\frac{1}{4}h^{i(2)''}{_{j}} + \frac{1}{2}\frac{a'}{a}h^{i(2)'}{_{j}} - \frac{1}{4}\nabla^{2}h^{i(2)}{_{j}}  + \partial^{i}\firstphi\partial_{j}\firstphi + 2\firstphi\partial^{i}\partial_{j}\firstphi \right.\\
    & \left. - 2\firstpsi\partial^{i}\partial_{j}\firstphi - \partial_{j}\firstphi\partial^{i}\firstpsi - \partial^{i}\firstphi\partial_{j}\firstpsi  + 3\partial^{i}\firstpsi\partial_{j}\firstpsi \right. \\
    & \left. + 4\firstpsi\partial^{i}\partial_{j}\firstpsi + 2\frac{a'}{a}\partial^{i}\firstw\partial_{j}{\firstphi} + 4\frac{a'}{a}\firstphi\partial^{j}\partial_{i}\firstw  + \dfirstphi\partial^{i}\partial_{j}\firstw \right. \\
    & \left. + 2\firstphi\partial^{i}\partial_{j}\dfirstw + \nabla^{2}\firstw\partial^{i}\partial_{j}\firstw - \partial_{j}\partial^{k}\firstw\partial^{i}\partial_{k}\firstw \right. \\
    & \left. - 2\frac{a'}{a}\partial^{i}\firstpsi\partial_{j}\firstw - 2\frac{a'}{a}\partial^{i}\firstw\partial_{j}\firstpsi - \partial^{i}\dfirstpsi\partial_{j}\firstw \right. \\
    & \left. + \partial^{i}\firstw\partial_{j}\dfirstpsi  - \partial^{i}\firstpsi\partial_{j}\dfirstw - \partial^{i}\dfirstw\partial_{j}\firstpsi \right. \\
    & \left. - 2\firstpsi\partial^{i}\partial_{j}\dfirstw + \dfirstpsi\partial^{i}\partial_{j}\firstw - 4\frac{a'}{a}\firstpsi\partial^{i}\partial_{j}\firstw \right. \\
    & \left. - 2\frac{a'}{a}\firstphi D^{i}{_{j}}\dfirsth - \frac{1}{2}\dfirstphi D^{i}{_{j}}\dfirsth - \firstphi D^{i}{_{j}}\ddfirsth \right. \\
    & \left. + \frac{1}{2}\partial_{k}\firstphi\partial^{i}D^{k}{_{j}}\firsth + \frac{1}{2}\partial_{k}\firstphi\partial_{j}D^{ki}\firsth - \frac{1}{2}\partial_{k}\firstphi\partial^{k}D^{i}{_{j}}\firsth \right. \\
    & \left. + \partial_{j}\partial_{k}\firstphi D^{ki}\firsth + \frac{1}{2}\dfirstpsi D^{i}{_{j}}\dfirsth + \ddfirstpsi D^{i}{_{j}}\firsth \right. \\
    & \left. + 2\frac{a'}{a}\dfirstpsi D^{i}{_{j}}\firsth + \frac{1}{2}\partial_{k}\firstpsi\partial^{i}D^{k}{_{j}}\firsth  + 2\frac{a'}{a}\firstpsi D^{i}{_{j}}\dfirsth \right. \\
    & \left. + \firstpsi D^{i}{_{j}}\ddfirsth + \frac{1}{2}\partial_{k}\firstpsi\partial_{j}D^{ki}\firsth - \frac{3}{2}\partial_{k}\firstpsi\partial^{k}D^{i}{_{j}}\firsth \right. \\
    & \left. + 2\firstpsi\partial_{k}\partial^{i}D^{k}{_{j}}\firsth + 2\firstpsi\partial_{k}\partial_{j}D^{ki}\firsth - 2\firstpsi\nabla^{2}D^{i}{_{j}}\firsth \right. \\
    & \left. - \nabla^{2}\firstpsi D^{i}{_{j}}\firsth + \partial^{i}\firstpsi\partial_{k}D^{k}{_{j}}\firsth + \partial_{j}\firstpsi\partial_{k}D^{ki}\firsth \right. \\
    & \left. + \partial_{k}\partial^{i}\firstpsi D^{k}{_{j}}\firsth + \frac{1}{2}\partial^{i}\firstw\partial_{k}D^{k}{_{j}}\dfirsth  + \frac{1}{2}\partial_{k}\partial^{i}\firstw D^{k}{_{j}}\dfirsth \right. \\
    & \left. + \frac{1}{2}\partial_{k}\partial_{j}\firstw D^{ki}\dfirsth - \frac{1}{2}\nabla^{2}\firstw D^{i}{_{j}}\dfirsth + \frac{1}{2}\partial^{k}\firstw\partial^{i}D_{kj}\dfirsth \right. \\
    & \left. + \frac{1}{2}\partial^{k}\firstw\partial_{j}D^{i}{_{k}}\dfirsth - \partial^{k}\firstw\partial_{k}D^{i}{_{j}}\dfirsth + \frac{1}{2}\partial^{k}\dfirstw\partial^{i}D_{kj}\firsth \right. \\
    & \left. + \frac{1}{2}\partial^{k}\dfirstw\partial_{j}D^{i}{_{k}}\firsth - \frac{1}{2}\partial^{k}\dfirstw\partial_{k}D^{i}{_{j}}\firsth + \partial_{k}\partial_{j}\dfirstw D^{ik}\firsth \right. \\
    & \left. + \frac{a'}{a}\partial^{k}\firstw\partial^{i}D_{kj}\firsth + \frac{a'}{a}\partial^{k}\firstw\partial_{j}D^{i}{_{k}}\firsth  - \frac{a'}{a}\partial^{k}\firstw\partial_{k}D^{i}{_{j}}\firsth \right. \\
    & \left. + 2\frac{a'}{a}\partial_{k}\partial_{j}\firstw D^{ik}\firsth - \frac{1}{2}D^{ki}\dfirsth D_{kj}\dfirsth  - \frac{1}{2}\partial^{i}D_{mj}\firsth\partial_{k}D^{km}\firsth \right. \\
    & \left. - \frac{1}{2}\partial_{j}D^{i}{_{m}}\firsth\partial_{k}D^{km}\firsth + \frac{1}{2}\partial_{m}D^{i}{_{j}}\firsth\partial_{k}D^{km}\firsth \right. \\
    & \left. - \frac{1}{2}\partial_{k}\partial_{j}D^{i}{_{m}}\firsth D^{km}\firsth  + \frac{1}{2}\partial_{k}\partial_{m}D^{i}{_{j}}\firsth D^{km}\firsth \right. \\
    & \left. + \frac{1}{2}D^{km}\firsth\partial^{i}\partial_{j}D_{km}\firsth   - \frac{1}{2}\partial_{k}\partial^{i}D_{mj}\firsth D^{km}\firsth \right. \\
    & \left.+ \frac{1}{4}\partial^{i}D^{mk}\firsth\partial_{j}D_{mk}\firsth - \frac{a'}{a}D_{kj}\dfirsth D^{ik}\firsth - \frac{1}{2}D_{kj}\ddfirsth D^{ki}\firsth \right. \\
    & \left.- \partial_{m}\partial_{k}D^{m}{_{j}}\firsth D^{ki}\firsth + \frac{1}{2}\partial_{m}\partial^{m}D_{kj}\firsth D^{ki}\firsth \right. \\
    & \left. + \frac{1}{2}\partial_{m}D^{ik}\firsth\partial^{m}D_{kj}\firsth - \frac{1}{2}\partial_{m}D^{ik}\firsth\partial_{k}D^{m}{_{j}}\firsth \right], \numberthis  
\end{align*}
\begin{align*}
\label{second_order_T}
    P^{li}_{jm}T^{(2)m}{_{l}} = & \frac{4}{a^{4}\kappa^{4}\left(\bar{\rho} + \bar{P}\right)}\left[\partial^{i}\left(\dfirstpsi + \mathcal{H}\firstphi\right)\partial_{j}\left(\dfirstpsi + \mathcal{H}\firstphi\right)\right] \\
    & + \frac{1}{a^{4}\kappa^{4}\left(\bar{\rho} + \bar{P}\right)}\left[\partial^{i}\left(\dfirstpsi + \mathcal{H}\firstphi\right)\partial_{k}D^{k}{_{j}}\dfirsth \right. \\
    & \left. + \partial_{j}\left(\dfirstpsi + \mathcal{H}\firstphi\right)\partial_{k}D^{ik}\dfirsth\right] \\
    & + \frac{1}{2a^{4}\kappa^{4}\left(\bar{\rho} + \bar{P}\right)}\partial_{k}D^{ik}\dfirsth\partial_{k}D^{k}{_{j}}\dfirsth \\
    & + \frac{2}{a^{2}\kappa^{2}}\left[\partial^{i}\firstw\mathcal{H}\partial_{j}\firstphi + \partial^{i}\firstw\partial_{j}\dfirstpsi - \frac{1}{4}\partial^{i}\firstw\partial_{k}D^{k}{_{j}}\dfirsth\right] \\
    & - \frac{1}{a^{2}\kappa^{2}}\left(4\firstpsi\delta^{ik} - 2D^{ik}\firsth\right)\left[\left(\partial_{j}\partial_{k} - \frac{1}{3}\nabla^{2}\delta_{jk}\right)\left(\firstphi - \firstpsi \right. \right. \\
    & \left. \left. + 2\mathcal{H}\firstw + \dfirstw\right)\right]. \numberthis
\end{align*}

\section{Source term of SIGW}\label{source_term}

The source term of the SIGW in a generic gauge is given by, 
\begin{align*}
\label{source_term_generic}
    S^{i}_{_{j}} =  & \left. \partial^{i}\firstphi\partial_{j}\firstphi + 2\firstphi\partial^{i}\partial_{j}\firstphi - 2\firstpsi\partial^{i}\partial_{j}\firstphi - \partial_{j}\firstphi\partial^{i}\firstpsi - \partial^{i}\firstphi\partial_{j}\firstpsi \right. \\
    & \left. + 3\partial^{i}\firstpsi\partial_{j}\firstpsi + 4\firstpsi\partial^{i}\partial_{j}\firstpsi + 2\mathcal{H}\partial^{i}\firstw\partial_{j}{\firstphi} + 4\mathcal{H}\firstphi\partial^{j}\partial_{i}\firstw \right. \\
    & \left. + \dfirstphi\partial^{i}\partial_{j}\firstw + 2\firstphi\partial^{i}\partial_{j}\dfirstw + \nabla^{2}\firstw\partial^{i}\partial_{j}\firstw - \partial_{j}\partial^{k}\firstw\partial^{i}\partial_{k}\firstw \right. \\
    & \left. - 2\mathcal{H}\partial^{i}\firstpsi\partial_{j}\firstw - 2\mathcal{H}\partial^{i}\firstw\partial_{j}\firstpsi - \partial^{i}\dfirstpsi\partial_{j}\firstw + \partial^{i}\firstw\partial_{j}\dfirstpsi \right. \\
    & \left. - \partial^{i}\firstpsi\partial_{j}\dfirstw - \partial^{i}\dfirstw\partial_{j}\firstpsi - 2\firstpsi\partial^{i}\partial_{j}\dfirstw + \dfirstpsi\partial^{i}\partial_{j}\firstw \right. \\
    & \left. - 4\mathcal{H}\firstpsi\partial^{i}\partial_{j}\firstw - 2\mathcal{H}\firstphi D^{i}{_{j}}\dfirsth - \frac{1}{2}\dfirstphi D^{i}{_{j}}\dfirsth - \firstphi D^{i}{_{j}}\ddfirsth \right. \\
    & \left. + \frac{1}{2}\partial_{k}\firstphi\partial^{i}D^{k}{_{j}}\firsth + \frac{1}{2}\partial_{k}\firstphi\partial_{j}D^{ki}\firsth - \frac{1}{2}\partial_{k}\firstphi\partial^{k}D^{i}{_{j}}\firsth + \partial_{j}\partial_{k}\firstphi D^{ki}\firsth \right. \\
    & \left. \frac{1}{2}\dfirstpsi D^{i}{_{j}}\dfirsth + \ddfirstpsi D^{i}{_{j}}\firsth + 2\mathcal{H}\dfirstpsi D^{i}{_{j}}\firsth + \frac{1}{2}\partial_{k}\firstpsi\partial^{i}D^{k}{_{j}}\firsth \right. \\
    & \left. + 2\mathcal{H}\firstpsi D^{i}{_{j}}\dfirsth + \firstpsi D^{i}{_{j}}\ddfirsth + \frac{1}{2}\partial_{k}\firstpsi\partial_{j}D^{ki}\firsth - \frac{3}{2}\partial_{k}\firstpsi\partial^{k}D^{i}{_{j}}\firsth \right. \\
    & \left. + 2\firstpsi\partial_{k}\partial^{i}D^{k}{_{j}}\firsth + 2\firstpsi\partial_{k}\partial_{j}D^{ki}\firsth - 2\firstpsi\nabla^{2}D^{i}{_{j}}\firsth - \nabla^{2}\firstpsi D^{i}{_{j}}\firsth \right. \\
    & \left. + \partial^{i}\firstpsi\partial_{k}D^{k}{_{j}}\firsth + \partial_{j}\firstpsi\partial_{k}D^{ki}\firsth + \partial_{k}\partial^{i}\firstpsi D^{k}{_{j}}\firsth + \frac{1}{2}\partial^{i}\firstw\partial_{k}D^{k}{_{j}}\dfirsth \right. \\
    & \left. + \frac{1}{2}\partial_{k}\partial^{i}\firstw D^{k}{_{j}}\dfirsth + \frac{1}{2}\partial_{k}\partial_{j}\firstw D^{ki}\dfirsth - \frac{1}{2}\nabla^{2}\firstw D^{i}{_{j}}\dfirsth \right. \\
    & \left. + \frac{1}{2}\partial^{k}\firstw\partial^{i}D_{kj}\dfirsth  + \frac{1}{2}\partial^{k}\firstw\partial_{j}D^{i}{_{k}}\dfirsth - \partial^{k}\firstw\partial_{k}D^{i}{_{j}}\dfirsth \right. \\
    & \left. + \frac{1}{2}\partial^{k}\dfirstw\partial^{i}D_{kj}\firsth + \frac{1}{2}\partial^{k}\dfirstw\partial_{j}D^{i}{_{k}}\firsth  - \frac{1}{2}\partial^{k}\dfirstw\partial_{k}D^{i}{_{j}}\firsth \right. \\
    & \left. + \partial_{k}\partial_{j}\dfirstw D^{ik}\firsth + \mathcal{H}\partial^{k}\firstw\partial^{i}D_{kj}\firsth + \mathcal{H}\partial^{k}\firstw\partial_{j}D^{i}{_{k}}\firsth \right. \\
    & \left. - \mathcal{H}\partial^{k}\firstw\partial_{k}D^{i}{_{j}}\firsth + 2\mathcal{H}\partial_{k}\partial_{j}\firstw D^{ik}\firsth - \frac{1}{2}D^{ki}\dfirsth D_{kj}\dfirsth \right. \\
    & \left. - \frac{1}{2}\partial^{i}D_{mj}\firsth\partial_{k}D^{km}\firsth - \frac{1}{2}\partial_{j}D^{i}{_{m}}\firsth\partial_{k}D^{km}\firsth + \frac{1}{2}\partial_{m}D^{i}{_{j}}\firsth\partial_{k}D^{km}\firsth \right. \\
    & \left. - \frac{1}{2}\partial_{k}\partial^{i}D_{mj}\firsth D^{km}\firsth - \frac{1}{2}\partial_{k}\partial_{j}D^{i}{_{m}}\firsth D^{km}\firsth + \frac{1}{2}\partial_{k}\partial_{m}D^{i}{_{j}}\firsth D^{km}\firsth \right. \\
    & \left. + \frac{1}{2}D^{km}\firsth\partial^{i}\partial_{j}D_{km}\firsth + \frac{1}{4}\partial^{i}D^{mk}\firsth\partial_{j}D_{mk}\firsth \right. \\
    & \left. - \mathcal{H}D_{kj}\dfirsth D^{ik}\firsth - \frac{1}{2}D_{kj}\ddfirsth D^{ki}\firsth - \partial_{m}\partial_{k}D^{m}{_{j}}\firsth D^{ki}\firsth \right. \\
    & \left. + \frac{1}{2}\partial_{m}\partial^{m}D_{kj}\firsth D^{ki}\firsth + \frac{1}{2}\partial_{m}D^{ik}\firsth\partial^{m}D_{kj}\firsth - \frac{1}{2}\partial_{m}D^{ik}\firsth\partial_{k}D^{m}{_{j}}\firsth \right.\\
    & - \frac{4}{3\mathcal{H}^{2}\left(1 + w\right)}\left[\partial^{i}\left(\dfirstpsi  + \mathcal{H}\firstphi\right)\partial_{j}\left(\dfirstpsi  + \mathcal{H}\firstphi\right)\right] \\
    & - \frac{1}{3\mathcal{H}^{2}\left(1 + w\right)}\left[\partial^{i}\left(\dfirstpsi  + \mathcal{H}\firstphi\right)\partial_{k}D^{k}{_{j}}\dfirsth  \right. \\
    & \left. + \partial_{j}\left(\dfirstpsi  + \mathcal{H}\firstphi\right)\partial_{k}D^{ki}\dfirsth \right] \\
    & - \frac{1}{6\mathcal{H}^{2}\left(1 + w\right)}\partial_{k}D^{ki}\dfirsth \partial_{k}D^{k}{_{j}}\dfirsth  \\
    & - 2\left[\partial^{i}\firstw\mathcal{H}\partial_{j}\firstphi + \partial^{i}\firstw\partial_{j}\dfirstpsi  + \frac{1}{4}\partial^{i}\firstw\partial_{k}D^{k}{_{j}}\dfirsth \right] \\
    & + \left(4\firstpsi\delta^{ik} - 2D^{ik}\firsth\right)\left[\left(\partial_{j}\partial_{k} - \frac{1}{3}\nabla^{2}\delta_{jk}\right)\left(\firstphi - \firstpsi \right. \right. \\
    & \left. \left.+ 2\mathcal{H}\firstw + \dfirstw \right)\right]. \numberthis
\end{align*}

\section{\label{subsec:analytical_kernel}Analytical expression of the source function}

Here we present the analytical expression of the kernel in all three gauges. Starting with the synchronous gauge. We substitute the transfer functions \cref{first_order_solution_synchronous_psi}--\cref{first_order_solution_synchronous_sigma} into \cref{sourceterm_sync_gr_simplified} to find
\begin{align*}
\label{analytical_source_sync}
    f_{S}(u,v,\tau) = \frac{2}{u^3 v^3 x^5} & \left(-9 u v x \left(x^2 \left(u^2-v^2+1\right)+8 u+4 v-10\right) \right. \\
    & \left. + 3 \sin \left(\frac{v x}{\sqrt{3}}\right) \left(\sqrt{3} u \left(x^2 \left(3 \left(u^2+1\right)+(4 u-7) v^2\right)+12 v\right) \right. \right. \\
    & \left. \left. +x \left(u^2 \left(2 v \left(v x^2-3\right)-9\right)-12 u v^2+9 v^2-9\right) \sin \left(\frac{u x}{\sqrt{3}}\right)\right) \right. \\
    & \left. +3 \sqrt{3} v \left(x^2 \left(u^2 (2 v-1)-3 v^2+3\right)+24 u\right) \sin \left(\frac{u x}{\sqrt{3}}\right) \right. \\
    & \left. +6 u v \left(2 \sqrt{3} \left(u x^2-6\right) \sin \left(\frac{u x}{\sqrt{3}}\right)+3 (4 u-5) x\right) \cos \left(\frac{v x}{\sqrt{3}}\right) \right. \\
    & \left. +6 u v \cos \left(\frac{u x}{\sqrt{3}}\right) \left(2 \sqrt{3} \left(v x^2-3\right) \sin \left(\frac{v x}{\sqrt{3}}\right) \right. \right. \\
    & \left. \left. +3 (2 v-5) x+15 x \cos \left(\frac{v x}{\sqrt{3}}\right)\right)\right). \numberthis
\end{align*}
The source function for Poisson and uniform curvature gauge are found by substituting \cref{approximate_solution_phi_radiation} into \cref{sourceterm_poisson_gr_simplified} and \cref{sourceterm_spatially_flat_simplified} respectively,
\begin{align*}
\label{analytical_source_poisson}
    f_{P}(u,v,\tau) = \frac{12}{u^3 v^3 x^6} & \left(\sin \left(\frac{u x}{\sqrt{3}}\right) \left(\left(u^2 x^2 \left(v^2 x^2-6\right)-6 v^2 x^2+54\right) \sin \left(\frac{v x}{\sqrt{3}}\right)  \right. \right. \\
    & \left. \left. + 2 \sqrt{3} v x \left(u^2 x^2-9\right) \cos \left(\frac{v x}{\sqrt{3}}\right)\right) \right. \\
    & \left. +2 u x \cos \left(\frac{u x}{\sqrt{3}}\right) \left(\sqrt{3} \left(v^2 x^2-9\right) \sin \left(\frac{v x}{\sqrt{3}}\right)\right. \right. \\
    & \left. \left. +9 v x \cos \left(\frac{v x}{\sqrt{3}}\right)\right)\right), \numberthis
\end{align*}
\begin{align*}
\label{analytical_source_uniform}
    f_{U}(u,v,\tau) = -\frac{6}{u^3 v^3 x^4} & \left(\sin \left(\frac{u x}{\sqrt{3}}\right) \left(3 \left(-5 u^2+3 v^2+3\right) \sin \left(\frac{v x}{\sqrt{3}}\right) \right. \right. \\
    & \left. \left. +\sqrt{3} v x \left(5 u^2+5 v^2-3\right) \cos \left(\frac{v x}{\sqrt{3}}\right)\right)  \right. \\
    & \left. +u x \cos \left(\frac{u x}{\sqrt{3}}\right) \left(v x \left(3 u^2-5 v^2+3\right) \cos \left(\frac{v x}{\sqrt{3}}\right) \right. \right. \\
    & \left. \left. -3 \sqrt{3} \left(u^2+v^2+1\right) \sin \left(\frac{v x}{\sqrt{3}}\right)\right)\right). \numberthis
\end{align*}

\end{appendices}

\bibliography{SIGW_gauge_revised}% common bib file

\end{document}